Goran Petrevski

Ss. Cyril and Methodius University in Skopje, Faculty of Economics

Orchid ID: 0000-0002-5583-8287

E-mail: goran@eccf.ukim.edu.mk

# Macroeconomic Effects of Inflation Targeting: A Survey of the Empirical Literature

This paper surveys the voluminous empirical literature of inflation targeting (IT). Specifically, the paper focuses on three main issues: the main institutional, macroeconomic, and technical determinants that affect the adoption of IT; the effects of IT on macroeconomic performance (inflation expectations, inflation persistence, average inflation rate, inflation variability, output growth, output volatility, interest rates, exchange rates, and fiscal outcomes); and disinflation costs of IT (the so-called sacrifice ratios). The main findings from our review are the following: concerning the determinants behind the adoption of IT, there is robust empirical evidence that larger and more developed countries are more likely to adopt the IT regime; similarly, the introduction of this regime is conditional on previous disinflation, greater exchange rate flexibility, central bank independence, and higher level of financial development; however, the literature suggests that the link between various macroeconomic and institutional determinants and the likelihood of adopting IT may be rather weak, i.e., they are not to be viewed either as strict necessary or sufficient conditions; the empirical evidence has failed to provide convincing evidence that IT itself may serve as an effective tool for stabilizing inflation expectations and for reducing inflation persistence; the empirical research focused on advanced economies has failed to provide convincing evidence on the beneficial effects of IT on inflation performance, concluding that inflation targeters only converged towards the monetary policy of non-targeters, while there is some evidence that the gains from the IT regime may have been more prevalent in the emerging market economies (EMEs); there is not convincing evidence that IT is associated with either higher output growth or lower output variability; the empirical research suggests that IT may have differential effects on exchange-rate volatility in advanced economies versus EMEs; although the empirical evidence on the impact of IT on fiscal policy is quite limited, it supports the idea that IT indeed improves fiscal discipline; the empirical support to the proposition that IT is associated with lower disinflation costs seems to be rather weak. Therefore, the accumulated empirical literature implies that IT does not produce superior macroeconomic benefits in comparison with the alternative monetary strategies or, at most, they are quite modest.

Keywords: inflation targeting; monetary policy, inflation, emerging markets economies

JEL classification codes: E42, E52, E58

## 1. Introduction

In 1990, New Zealand adopted IT as a new framework for conducting monetary policy, followed by Canada (1991), United Kingdom (1992), and several other industrialized countries. Since then, this monetary policy regime has gained increasing popularity in both advanced countries and EMEs. As a result, by 2019, in one form or another, IT has been implemented in 43 countries as diverse as Albania, Ghana, Mexico, Russia, South Africa, Sweden, and UK (IMF, 2020). Commenting on the global popularity of IT, Rose (2007) claims that the international monetary system has been dominated by inflation targeters. Also, IT is the longest-lasting monetary strategy after the World War II. In addition, unlike the other monetary strategies, IT has proved to be durable as no country has left it, yet. Therefore, according to Walsh (2009), the actual experience with IT unambiguously shows that it is both feasible and sustainable.

Table 1 in the Appendix presents the IT adoption dates for selected industrialized countries and EMEs. As can be seen, in several cases it is quite difficult to specify the exact date of adopting the IT regime, mainly due to the variations in its practical implementation. For instance, Chile introduced IT in 1990-1991, while retaining its exchange rate band by August 1999, when it switched to full-fledged IT; similarly, Israel adopted explicit inflation target in 1992, but retained the exchange rate band through 1997; Mexico, too, introduced some elements of IT in 1999 though it had not moved to the full-fledged variant until 2002. In these regards, not only the adoption dates differ among the individual empirical studies, but there are also discrepancies between the dates that can be found in the empirical literature and those specified on the central bank's web sites and in the official documents.

According to Bernanke and Mishkin (1997), Hammond (2012), Mishkin (2000), Mishkin and Posen (1997), and Svensson (2002, 2010), practical implementation of IT is characterized by the following common features: strong institutional commitment to price stability as the primary monetary policy objective in the medium-to-long run; the announcements of explicit numerical inflation targets for medium-term inflation; short-term flexibility, which allows the policy makers to respond to short-term disturbances from various sources (supply shocks, exchange rate changes, etc.); as well as high degree of central bank independence, accountability, and transparency. As for the practical implementation of monetary policy, a distinctive feature of IT is the absence of intermediate targets, which stands in sharp contrast to the alternative monetary policy strategies, such as monetary or exchange rate targeting. In these regards, Bernanke and Mishkin (1997) describes IT as a "rule-like strategy" or "constrained discretion", which enables the central bank to be focused on price stability while at the same time being able to deal with short-run macroeconomic fluctuations.

Consequently, IT is supposed to provide the following advantages in comparison with other monetary policy regimes: it builds discipline, credibility and accountability of central banks by preventing policy makers to engage in systematic short-term stimuli, and by subjecting the central bank's short-run actions to public scrutiny and debate about their long-term consequences; it improves central bank's communication with the general public; it is both efficient and forward-looking strategy as it use all the available information along with an explicit account of time-lags; it helps the central bank to anchor inflation expectations and to cope with adverse supply shocks, which results in lower economic costs (Batini and Laxton 2007, Bernanke and Mishkin 1997, Mishkin 2000). Similarly, Mishkin and Schmidt-Hebbel (2002) argue that IT offers several benefits for EMEs, such as reinforcing central bank independence and enabling central banks to be more focused on inflation. In addition, Thornton and Vasilakis (2017) show that IT facilitates the implementation of countercyclical monetary policy in these countries, majority of which have been previously notorious for implementing procyclical policies.

The proponents of IT often emphasize its flexibility as a crucial property in the practical implementation of monetary policy (Mishkin 1999 and 2004a). On the one hand, the firm focus on price stability increases the credibility of central banks with its favourable effects on inflation expectations. On the other hand, central banks typically approach the inflation target gradually over time, thus, retaining the manoeuvre room for responding to possible adverse short-run circumstances. In other words, within this policy framework, central banks can combine the inflation targets with other policy goals such as output or employment (Agénor 2002, Leiderman and Svensson 1995, Svensson 1997a). In this way, flexible IT appears to be an optimal monetary policy leading to lower average inflation accompanied by output stabilization (Ball 1999a and 1999b, Clarida et al. 1999, Svensson 1997b).

On the other hand, IT has been criticizes on various grounds: a) it is too rigid by constraining discretion in monetary policy, thus, unnecessarily restraining growth; b) it cannot anchor inflation expectations because it offers too much discretion with respect to both the definition and maintenance of inflation targets; c) relatively frequent misses of inflation targets, due to the imperfect control of inflation and the long lags in the monetary transmission mechanism, can lead to weak central bank credibility; d) it may not be sufficient to ensure fiscal discipline or prevent fiscal dominance; e) the exchange rate flexibility required by IT might cause financial instability; and f) its practical implementation is dependent on a number of institutional and technical preconditions, which are not met in the most of EMEs (Batini and Laxton 2007, Bernanke et al. 1999, Mishkin 1999 and 2000). The last four disadvantages are especially relevant for EMEs. IT has been implemented in EMEs within specific macroeconomic and institutional environment, which undoubtedly affects the implementation of effective monetary policy. For instance, most of these countries are characterised by fiscal dominance and weak banking systems, which are not consistent with a sustainable IT regime. Also, as the long history of high inflation undermines the central banks' credibility, the introduction of IT should be preceded by, at least partial, disinflation. Finally, simultaneously with inflation targets, central banks in EMEs should take care for smoothing excessive exchange rate fluctuations for at least two reasons: the exchange rate channel is of crucial importance in small open economies; both dollarization and the exposure of sudden stops of foreign capital amplifies the vulnerability of firms and banks to exchange rate fluctuations, which may lead to full-blown financial crisis. Hence, central banks in EMEs must be concerned with exchange rate fluctuations, thus preventing sharp depreciations that might cause high inflation and financial instability. Yet, putting too much emphasis on the exchange rates might create confusion in the public, thus, compromising the credibility of inflation targets. Therefore, the practical implementation of IT in EMEs can be quite complicated: on the one hand, central banks should smooth exchange rate fluctuations, but on the other hand, they cannot allow the inflation targets to be subordinated to exchange rate policy.[1] Given these unfavourable macroeconomic and institutional conditions prevailing in EMEs, IT need not necessarily provide the outcomes that are either *a priori* expected in theory or observed in advanced economies. At the same time, as suggested by Walsh (2009), the larger variation in inflation experiences in EMEs may help identify the true effects of IT. Consequently, a large body of empirical evidence on the macroeconomic effects of IT has focused on EMEs.

This main goal of this article is to survey and to synthesize the main findings in the empirical literature on the implementation of IT. Specifically, the paper focuses on three main issues: the institutional, macroeconomic, and technical determinants that affect the adoption of IT (especially in EMEs); the effects of IT on macroeconomic performance (inflation expectations, average inflation, inflation variability, output growth, output volatility, interest rates, exchange rates etc.); and the effectiveness of IT as a tool for reducing inflation, i.e., whether it is associated with lower disinflation costs (the so-called sacrifice ratio). Despite the accumulated empirical literature in this field, it is quite surprising to observe the lack of comprehensive and well-structured review of it. To our best knowledge, only Angeriz and Arestis (2007b) summarize the available empirical literature on macroeconomic effects of IT. Hence, we believe that our paper provides a valuable contribution in filling this void. However, note that this survey does not cover other relevant fields in the empirical literature, such as the effects of IT on financial stability (Armand 2017, Fazio et al. 2018, Fouejieu 2017, Gong and Qian 2022, Kim and Mehrotra 2017), financial markets (O'Sullivan and Tomljanovich 2012), and international trade (McCloud and Taylor 2022).

The main findings from our review are the following: concerning the determinants behind the adoption of IT, there is robust empirical evidence that larger and more developed countries are more likely to adopt the IT regime; similarly, the introduction of this regime is conditional on previous disinflation, greater exchange rate flexibility, central bank independence, and higher level of financial development; however, the literature suggests that the link between various macroeconomic and institutional determinants and the likelihood of adopting IT may be rather weak, i.e., they are not to be viewed either as strict necessary or sufficient conditions; the empirical evidence has failed to provide convincing evidence that IT itself may serve as an effective tool for stabilizing inflation expectations and for reducing inflation persistence; the empirical research focused on advanced economies has failed to provide convincing evidence on the beneficial effects of IT on inflation performance, concluding that inflation targeters only converged towards the monetary policy of non-targeters, while there is some evidence that the gains from the IT regime may have been more prevalent in the EMEs; there is not convincing evidence that IT is associated with either higher output growth or lower output variability; the empirical research suggests that IT may have differential effects on exchange-rate volatility in advanced economies versus EMEs; although the empirical evidence on the impact of IT on fiscal policy is quite limited, it supports the idea that IT indeed improves fiscal discipline; the empirical support to the proposition that IT is associated with lower disinflation costs seems to be rather weak. Therefore, the accumulated empirical literature implies that IT does not produce superior macroeconomic benefits in comparison with the alternative monetary strategies or, at most, they are quite modest. That said, it seems that the increasing popularity of IT is not based on strong empirical evidence with respect to the macroeconomic performance of this regime.

[1] For theoretical discussion as well as empirical evidence on the importance of exchange rates for the adoption and implementation of IT, see: Agénor (2002), Buffie et al. (2018), Aizenman et al. (2010), Carrare et al. (2002), Céspedes et al. (2014), Civcir and Akçağlayan (2010), Edwards (2007), Eichengreen et al. (1999), Ho and McCauley (2003), Kumhof et al. (2007), Leiderman et al. (2006), Jonas and Mishkin (2007), Mishkin (2000, 2004b), Mishkin and Schmidt-Hebbel (2002), Parrado (2004), Roger et al. (2009), Schaechter et al. (2000), Schmidt-Hebbel and Werner (2002), Siregar and Goo (2010), Stone et al. (2009).

The rest of the paper is structured as follows: Section 2 reviews the relevant literature dealing with the main determinants behind the adoption of IT with a special focus on EMEs; in Section 3 we provide a comprehensive review of the empirical literature on the macroeconomic effects of IT; Section 4 presents the main methodological issues and findings regarding the impact of IT on disinflation costs; Section 5 summarizes the main conclusions.

## 2. Main determinants of the choice of IT

Although an increasing number of countries have adopted IT during the past two decades, many more of them, especially the EMEs, still rely on other strategies for controlling inflation. This naturally raises the question of which factors determine the choice of IT vis-a-vis alternative monetary policy regimes. In principle, this choice should be based on both theoretic grounds and empirical evidence. Theoretically, the choice of optimal monetary policy has been analysed within a well specified (usually, a small-scale) macroeconomic model by comparing the central bank's loss function under alternative policy rules. Here, a number of papers demonstrate that IT outperforms the alternative monetary policy rules in terms of inflation/output variability (Ball 1999a, 1999b, Haldane and Batini 1999, Rudebusch and Svensson 1999, Svensson 1999a, 1999b, 2000). At the same time, despite the accumulated empirical evidence on the macroeconomic effects of IT, the findings from these studies are rather inconclusive: while some papers suggest that IT is associated with both lower average inflation and improved inflation/output variability, others show that it does not produce superior macroeconomic benefits or, at most, they are quite modest (See the empirical evidence reviewed in Section 3). Therefore, the increasing adoption of IT is not based on strong empirical evidence with respect to the macroeconomic performance of this monetary regime. In addition, it should be noted that the experience of advanced countries may not be relevant for EMEs due to their specific institutional and macroeconomic characteristics.

The early literature has suggested that the adoption of IT requires the fulfilment of several economic, institutional and technical prerequisites, such as: the absence of fiscal dominance, strong external position, relatively low inflation, well-developed financial markets and sound financial system, central bank independence, some structural characteristics (price deregulation, low dollarization, low sensitivity to supply shocks, strong external position etc.), the absence of de facto exchange rate targets, well developed technical infrastructure for forecasting inflation etc. (Agénor 2002, Amato and Gerlach 2002, Battini and Laxton 2007, Carare et al. 2002, Carare and Stone 2006, Eichengreen et al. 1999, Freedman and Ötker-Robe 2009 and 2010, IMF 2006, Masson et al. 1997, Mishkin 2000, Mishkin and Savastano 2002, Mishkin and Schmidt-Hebbel 2002, Mishkin and Schmidt-Hebbel 2007).

While theoretically sound, the experience shows that many inflation targeters, especially the EMEs, have not met all these requirements, at least in the initial phase. Indeed, EMEs operate in a specific institutional and macroeconomic environment, which often complicates the design and implementation of the IT regime. For instance, the presence of fiscal dominance, a common feature in many EMEs, undermines the effectiveness of monetary policy. Similarly, weak banking systems in these countries often precludes the use of market-based monetary policy instruments.

Further on, the long historical experience with high inflation reduces the credibility of their central banks, requiring at least partial disinflation before the introduction of IT. Also, given the crucial importance of the exchange rate channel in small open economies, the central banks in EMEs must be concerned with both exchange rate fluctuations and inflation targets simultaneously. Finally, central banks in many EMEs often lack the necessary technical infrastructure (data availability, lack of systematic forecasting process, low understanding of the transmission mechanism etc.), which hampers the day-to-day implementation of IT (Amato and Gerlach 2002, Jonas and Mishkin 2007, Masson et al. 1997, Mishkin 2000 and 2004, Mishkin and Savastano 2002, Mishkin and Schmidt-Hebbel 2002).

In this regard, Masson et al. (1997) assess the monetary policy framework in five EMEs and show that they lag substantially by those prevailing among the inflation targeters. Therefore, they conclude that developing countries do not fulfil the requirements for adopting IT. Similarly, based on a survey of 31 central banks, Batini and Laxton (2007) assess whether some preconditions must be met before adopting IT in EMEs, such as: technical infrastructure, financial system, institutional central bank independence, and economic structure. They construct an extensive list of parameters and, by quantifying each of them, conclude that EMEs had not satisfied these required preconditions, which implies that adopting IT does not depend on meeting some strict initial pre-conditions. Similarly, based on the experience with the introduction and implementation of IT, Freedman and Ötker-Robe (2009) and Schmidt-Hebbel and Carrasco (2016) argue that a country must meet some basic preconditions before adopting IT, though most of the countries failed to meet all the preconditions. More importantly, they show that the adoption of IT itself promotes the fulfilment of these preconditions. Samarina and Sturm (2014) provide a strong empirical support to this hypothesis by showing that there is a structural change after the adoption of IT, implying that, even when a country does not meet all the preconditions, once it has adopted IT, this decision leads to changes in the institutions which support its proper functioning.

The empirical literature on the determinants behind the adoption of IT generally follows an eclectic approach by specifying a general list of determinants that are expected to affect the choice of IT. In other words, only a few studies focus on the role of specific factors (e.g., political). Consequently, this approach prevents us to provide a structured survey of this strand of literature. In addition to the empirical studies investigating explicitly the determinants of IT, there are number of papers which, although having different research topic, deal with issue as part of the overall empirical approach. Here, we refer to the papers employing the propensity score matching methodology or other similar types of treatment effects regression. Within this framework, in the first stage of the empirical investigation the dummy variable of adopting IT is usually regressed on several macroeconomic variables. The non-exhaustive list of this research includes Ardakani et al. (2018), Arsić et al. (2022), de Mendonça and de Guimarães e Souza (2012), Fry-McKibbin and Wang (2014), Gonçalves and Carvalho (2009), Lin (2010), Lin and Ye (2007, 2009), Lucotte (2012), Minea and Tapsoba (2014), Minea et al. (2021), Mukherjee and Singer (2008), Pontines (2013), Samarina et al. (2014), Vega and Winkelried (2005), and Yamada (2013).

In what follows we first provide a brief explanation of the expected impact of the above-mentioned determinants on the likelihood of adopting IT along with an overview of the main findings from the empirical research in this field. Table 2A in the Appendix provides detailed description of individual studies, while Table 2B summarizes the main empirical findings by each determinant.

Besides the well-known argument that small open economies are the most serious candidates for pegged exchange rates, *a priori*, it is difficult to say whether IT is a "one-size-fits-all" strategy which is appropriate for both large and small economies. At the same time, the experience reveals that IT has been implemented in a wide array of countries, ranging from very small (Albania, Israel, Serbia etc.) to very large countries (Brazil, India, Russia, and South Africa). Most of the empirical studies confirm that size matters for the adoption of IT by showing that the size of the economy, measured either by the level of GDP or GDP per capita, is associated with higher likelihood of adopting this monetary regime (de Mendonça and de Guimarães e Souza 2012, Leyva 2008, Lucotte 2010, Minea et al. 2021, Samarina et al. 2014, Yamada 2013). Yet, since GDP per capita is used as a general proxy for the level of economic development, some of these results imply that not larger but more developed economies are more likely to adopt IT. In a similar fashion, several studies obtain the same findings working with either area or population size (Arsić et al. 2022, Rose 2014, Wang 2016, Yamada 2013). However, this conclusion is not shared by Hu (2006) and Ismailov et al. (2016), while some papers suggest that the importance of this determinants may be sensitive to the sample. For instance, Carare and Stone (2006) and Fouejieu (2017) are not able to confirm this hypothesis for the sample of EMEs, while Samarina et al. (2014) provide similar evidence for the advanced economies.

Usually, central banks tend to choose their monetary strategy in response to past macroeconomic performance. In theory, inflation and output are the standard elements in the central bank's loss function. In practice, although inflation control is the primary goal of monetary policy, central banks often pay attention to economic activity, too. In this regard, if a country has experienced unsatisfactory economic performance, such as low growth rates or high output volatility, then the central bank might consider switching to IT as a strategy which enables policy makers to focus on the developments in the real economy, too. This argument may be especially relevant for EMEs, which traditionally have worse performance than advanced economies due to the unfavourable macroeconomic environment prevailing in them (Fraga et al. 2003). The empirical literature offers mixed evidence on the effects of economic performance on the choice of IT. For instance, Lucotte (2012) finds that higher output growth increases the likelihood to adopt IT. On the other hand, a few papers find a negative association between GDP growth and the probability of adopting IT, implying that the countries experiencing satisfactory economic performance have less incentives to switch to this regime (Ardakani et al. 2018, Hu 2006). In fact, most of the empirical research has produced either statistically insignificant or non-robust results about the importance of output growth for the choice of IT (Fouejieu 2017, Lin 2010, Lin and Ye 2007, Lin and Ye 2009, Pontines 2013, Samarina and De Haan 2014, Thornton and Vasilakis 2017, Wang 2016). Similarly, the empirical literature does not provide an unambiguous answer to the question of whether the countries facing more (less) stable economic conditions (measured by output volatility) are good (bad) candidates to implement IT. Mukherjee and Singer (2008) as well as Samarina and De Haan (2014) show that higher output volatility makes adopting IT more likely, Hu (2006) finds that this factor is statistically insignificant, while Samarina and Sturm (2014) and Stojanovikj and Petrevski (2019) obtain opposite findings, finding that provide some evidence that macroeconomic instability reduces the likelihood of adopting IT.

Similarly, the central bank might choose its monetary policy strategy based on its experience with past inflation rate. Here, the literature suggests that the introduction of IT is not feasible at high inflation rates, when there is a considerable degree of inertia in nominal variables, and monetary policy is largely accommodative. Therefore, a country should first reduce inflation to a relatively low level before it adopts this monetary regime (Carare et al. 2002, Masson et al. 1997, Mishkin 2000). Accordingly, this argument implies that higher inflation rates make the introduction of IT less likely. The empirical research provides strong support to this proposition for both industrialized countries and EMEs (Ardakani et al. 2018, Arsić et al. 2022, Hu 2006, Lin 2010, Lin and Ye 2007, Lin and Ye 2009, Minea and Tapsoba 2014, Minea et al. 2021, Pontines 2013, Samarina and Sturm 2014, Thornton and Vasilakis 2017). Yet, it is fair to note that the empirical evidence is not unanimous: Gonçalves and Carvalho (2009) and Vega and Winkelried (2005) obtain opposite findings, while the results in Fry-McKibbin and Wang (2014), Samarina and De Haan (2014), Samarina et al. (2014), and Wang (2016) are either insignificant or sensitive to the sample they work with (industrialized countries versus EMEs). Given the consensus view of inflation as a monetary phenomenon in the long-run, several studies test the relationship between money growth and the probability of adopting IT. It seems that this consensus prevails in the empirical literature, too (Ardakani et al. 2018, Arsić et al. 2022, Lin 2010, Lin and Ye 2007, Lin and Ye 2009, Pontines 2013, Samarina et al. 2014, Yamada 2013), with only a few exceptions (Fry-McKibbin and Wang 2014, Wang 2016). Therefore, it is safe to say that the countries experiencing higher past inflation or, equivalently, higher money growth, are less likely to switch to IT.

Strong external position, too, is expected to make the adoption of IT more likely. Within this monetary regime, the central bank should be focused on achieving and maintaining the inflation targets, which is only possible if the concerns for the balance payment and the exchange rate are subordinated to the primary objective of monetary policy (Carare et al. 2002). However, the available empirical literature provides ambiguous findings on the importance of external macroeconomic conditions for the adoption of IT. For instance, Arsić et al. (2022) show that strong current account position reduces the likelihood of adopting IT, Mukherjee and Singer (2008) obtain opposite findings, while this variable is not significant in Hu (2006). The evidence on the role of external debt is equally inconclusive: Hu (2006) finds that higher external debt reduces the probability of adopting IT, while this factor is not significant in Samarina and De Haan (2014). Working with a sample of EMEs, Yamada (2013) finds that foreign exchange reserve, too, is not a significant factor when switching to IT. Similarly, Ardakani et al. (2018) obtain opposite results on the importance of central bank's assets: while their size makes the adoption of IT more likely in the advanced countries, it is quite contrary for the case of developing countries. In fact, all this evidence suggests that, although strong external position may make the transition toward IT easier, the developing countries characterized by favorable current account balance and/or sizable foreign exchange reserves have less incentives to change their existing monetary regimes (usually, some variant of a currency peg).

In addition, fiscal discipline is often listed as one of the basic requirements for adopting IT. In the presence of persistent high fiscal deficits, the central bank may be pursue accommodative monetary policy, which clearly undermines its ability to meet the announced inflation targets. Similarly, a high level of public debt may provide an incentive for the government to reduce the real value of the debt by high inflation (Mishkin 2000). As a result, fiscal discipline, and sound public finance in general (efficient tax-collection procedures, high government revenue, low budget deficits, and low public debt), are expected to increase the likelihood of adopting IT. However, the empirical evidence on the role of fiscal discipline is rather mixed, and this is equally true for the importance of both the budget balance and the public debt. As for role of budget balance, only a few studies support the above proposition (Hu 2006, Lin 2010), while the majority of the empirical research either rejects it (de Mendonça and de Guimarães e Souza 2012, Pontines 2013) or provides inconclusive evidence (Carare and Stone 2006, Leyva 2008, Lin and Ye 2007, Mishkin and Schmidt-Hebbel 2002, Samarina and De Haan 2014, Vega and Winkelried 2005). The empirical literature is equally inconclusive when employing government debt as a fiscal policy indicator. Here, only a few studies show that higher indebtedness reduces the probability of adopting IT (Gonçalves and Carvalho 2009, Minea and Tapsoba 2014, Thornton and Vasilakis 2017), while others obtain either opposite findings (Arsić et al. 2022) or provide mixed evidence (Carare and Stone 2006, Ismailov et al. 2016, Lin and Ye 2009, Samarina and De Haan 2014, Samarina and Sturm 2014, Samarina et al. 2014, Wang 2016). Therefore, the empirical evidence implies that the role of fiscal discipline in choosing IT may be conditional on other factors, such as a country's history with inflation, the government's access to financial markets, central bank independence, the limits on central bank lending to the government etc. For instance, in countries with a long history of low inflation and with broad markets for government debt, the credibility of IT is less dependent on the government's actual fiscal position. Also, central bank independence accompanied by clear limits on central bank lending to the government diminish the role of fiscal discipline in the decision-process (Carare et al. 2002). Therefore, the inconclusive evidence on the role of fiscal discipline for adopting IT may not be surprising for the case of industrialized countries, which are characterized by long history of low inflation, broad and deep markets for government debt, and strong institutional environment. However, the lack of firm evidence is puzzling for the case of developing countries despite their long record of fiscal dominance, high inflation, and low central bank credibility.

Both trade and financial openness of the economy are also considered relevant factors for the choice of monetary policy strategy. For instance, many EMEs are traditionally exposed to large and persistent exogenous shocks, which makes them very sensitive to commodity prices and exchange rate fluctuations (Fraga et al. 2003). Consequently, small open economies tend to choose currency pegs as a preferred monetary regime, thus, being less likely to switch to IT (IMF 2006, Rose 2014). As for the importance of trade openness, the empirical literature seems to be completely divided about the importance of this factor: while a few studies find that trade openness is associated with higher probability of adopting IT (Leyva 2008, Lucotte 2010, Mishkin and Schmidt-Hebbel 2002), the majority of empirical research fails to support this proposition (Arsić et al. 2022, de Mendonça and de Guimarães e Souza 2012, Fouejieu 2017, Fry-McKibbin and Wang 2014, Hu 2006, Lucotte 2012, Lin 2010, Lin and Ye 2007, Lin and Ye 2009, Minea et al. 2021, Minea and Tapsoba 2014, Rose 2014, Samarina and De Haan 2014, Samarina et al. 2014, Thornton and Vasilakis 2017, Vega and Winkelried 2005), and this is true for both advanced economies and EMEs. On the other hand, most of the empirical suggests that financial openness is an important precondition for adopting IT (de Mendonça and de Guimarães e Souza 2012, Samarina et al. 2014, Thornton and Vasilakis 2017) though a few studies refute this conclusion (Rose 2014, Samarina and De Haan 2014, Samarina et al. 2014).

Within the IT framework, price stability is the primary objective of monetary policy with other objectives (employment, exchange rate, external position) being subordinated to the inflation target. Therefore, by definition, IT requires flexible exchange rates, i.e., it is inconsistent with fixed exchange rate regimes. In other words, the presence of fixed exchange rates is expected to decrease the likelihood of adopting IT, while greater exchange rate flexibility works in the opposite direction. Indeed, the available empirical literature unanimously confirm that fixed exchange rates are not conducive to IT (Ardakani et al. 2018, Arsić et al. 2022, de Mendonça and de Guimarães e Souza 2012, Fouejieu 2017, Fry-McKibbin and Wang 2014, Lin 2010, Lin and Ye 2007, Lin and Ye 2009, Minea and Tapsoba 2014). Similarly, with a few exceptions (Hu 2006, Samarina and Sturm 2014), the large majority of empirical research finds that exchange rate flexibility makes the adoption of IT more likely (Ismailov et al. 2016, Lucotte 2010, Lucotte 2012, Minea et al. 2021, Mukherjee and Singer 2008, Pontines 2013, Samarina and De Haan 2014, Samarina et al.

2014, Thornton and Vasilakis 2017, Vega and Winkelried 2005). Therefore, we can conclude that there is a strong consensus that IT requires higher degree of exchange rate flexibility, i.e., currency pegs are not compatible with this monetary regime.

Undoubtedly, central bank independence appears to be one of the most important institutional factors necessary for successful implementation of IT. It is also understood that the central bank should have a clear mandate to pursue price stability with all the other objectives being subordinated to the inflation target (Agénor 2002). It seems that this proposition has found a widespread empirical support (Fouejieu 2017, Lin and Ye 2007, Lucotte 2010 and 2012, Minea and Tapsoba 2014) with only a few dissenting studies (Hu 2006, Lin and Ye 2009). Here, the literature refers to the so-called instrument independence, i.e., the autonomy of central bank in choosing its instruments to achieve the inflation targets. In one of the first attempts to address the factors behind the choice of IT, Mishkin and Schmidt-Hebbel (2002), who conduct a cross-section analysis on a sample of 27 advanced countries and EMEs during 1990s. The main findings from their study indicate that it is the type of central bank independence that matters for the choice of IT. Specifically, they find that legal central bank independence is not significant in the choice of IT. In addition, they show that the likelihood of adopting IT is positively associated with instrument independence, but goal independence has the opposite impact. Samarina and De Haan (2014), too, confirm the importance of instrument independence for the sample of developing countries, but they find that this type of central bank independence is not significant for advanced economies. Carare and Stone (2006) investigate one particular dimension of central bank independence – restrictions on government lending and find that it is important only for the EMEs. Finally, Mukherjee and Singer (2008) show that central banks with a clear focus on price stability, i.e., those without bank regulatory authority, are more likely to choose IT.

Financial development and financial stability facilitate the adoption of IT in many ways. A well-developed financial system not only enables the central bank to employ market-based instruments, but it has a central role in the monetary policy transmission mechanism. Also, operating in a sound financial system, the central bank is free from the responsibility to inject liquidity to the failing financial institutions, so that it can focus on the achievement of the announced inflation target (Battini and Laxton 2007, Carare et al. 2002). The available empirical evidence generally supports the proposition that higher level of financial development is required for introducing IT (Carare and Stone 2006, de Mendonça and de Guimarães e Souza 2012, Leyva 2008, Samarina and Sturm 2014, Samarina et al. 2014, Thornton and Vasilakis 2017, Vega and Winkelried 2005) although it is fair to say that the empirical support is far from unanimous (Ardakani et al. 2018, Hu 2006, Lucotte 2010, Lucotte 2012, Samarina and De Haan 2014). On the other hand, Samarina and De Haan 2014 find that financial structure (market-based versus bank-based financial systems) does not matter for adopting IT. Also, the empirical literature fails to provide a clear conclusion on the importance of financial (in)stability: Samarina and De Haan (2014) find that the financial crisis dummy is statistically insignificant, while Thornton and Vasilakis (2017) provide some weak evidence that this factor matters in developing countries only.

Further on, several studies focus on the importance of political institutions for the adoption of IT. For instance, comparing IT with exchange rate pegs for a large set of more than 170 countries, Rose (2014) show that IT is a preferred monetary regime for the countries with more developed democratic institutions. Mukherjee and Singer (2008) show that countries are more likely to adopt IT when the government and the central bank share the same preferences for tight monetary policy. Specifically, the combination of a right-leaning government and a central bank without bank regulatory authority is likely to be associated with the adoption of IT. Ismailov et al. (2016) find that political stability does not affect the choice of IT for both low-income and high-income countries. Lucotte (2010 and 2012) investigates the role of institutional and political factors in adopting IT for a sample of 30 EMEs. His findings imply that a number of political determinants increase the likelihood of adopting IT, such as the number of veto players in the political system, political stability as well as federalism (decentralization). In a similar fashion, working with a sample of 53 developing countries, Minea et al. (2021) obtain some other interesting results: on the one hand, they find that better institutional quality reduce the likelihood to switch to IT, while on the other hand, constraints on the executive makes the introduction of IT more likely. In this respect, the former finding seems to be at odds to their theoretical model linking the monetary regime, quality of institutions, and the sources of government finance, while the latter result conforms well to the predictions from the theoretical model.

Finally, there are some technical prerequisites for successful implementation of IT. For instance, the central bank should have a clear understanding of the time lag and the transmission mechanism; it should have long and reliable database and technical expertise to forecast inflation; it should conduct regular surveys of inflation expectations; and it should be able to develop market-based and forward-looking operating procedures. However, it is suggested that the initial technical conditions, although important, are not critical for introducing IT, i.e., the lack of these conditions can be remedied after the introduction of this monetary regime (Battini and Laxton 2007, and Carare et al. 2002, IMF 2006, Mishkin and Schmidt-Hebbel 2007).

Unsurprisingly, the empirical research has not led to firm conclusions on the importance of each individual determinant for the adoption of IT reflecting the fact that developed countries and EMEs represent a heterogeneous group with different institutional and macroeconomic characteristics. Indeed, several studies show that the determinants of the choice of IT generally differ between advanced economies and EMEs. For instance, Ardakani et al. (2018), Fry-McKibbin and Wang (2014), Ismailov et al. (2016), Samarina and De Haan (2014), Samarina et al. (2014), and Thornton and Vasilakis (2017) find that some macroeconomic variables are relevant in both the advanced and developing countries, whereas others may have differential impacts across these two groups of countries.

As suggested above, the importance of various determinants of IT may differ with respect to the type of IT regimes. For instance, Carare and Stone (2006) review the global experience with IT by focusing on the factors affecting the evolution between various variants of this regime ("lite", eclectic, and full-fledged). They find that the level of economic and financial development are the most significant factors for the overall central bank credibility and, thus, for the choice of IT regimes. Also, they discuss the experience of EMEs and show that the likelihood to move from "lite" to full-fledged IT is predominantly influenced by the level of financial development, government debt, and central bank restrictions on government financing. In their comprehensive study, Samarina and De Haan (2014), too, show that the most important factors behind the adoption of IT differ between soft and full-fledged inflation targeters. Specifically, they find that flexible exchange rate regimes, exchange rate volatility, central bank independence, and external debt affect the probability of adopting soft IT, whereas inflation, output growth, and public debt are the most important factors for adopting full-fledged IT.

Based on Table 2B in the Appendix, the main findings from the empirical literature can be summarized as follows: the empirical research generally suggests that larger economies are more likely to choose this monetary regime; also, the level of economic development is associated with higher likelihood of adopting IT; these findings imply that IT may not be a feasible monetary regime for small and/or low-income countries; however, the empirical literature offers diverse results on the effects of economic activity, i.e., it is not clear whether the adoption of IT is more likely in the countries with higher or lower output growth; similarly, the empirical literature does not provide unambiguous answer to the question of whether the countries facing more (less) stable economic conditions (measured by output volatility) are good (bad) candidates to introduce IT; on the other hand, there is strong empirical evidence that the countries experiencing higher past inflation or equivalently, higher money growth are less likely to switch to IT; this finding is consistent with the common requirement that the introduction of IT is conditional on previous disinflation; as for the external macroeconomic conditions, there are ambiguous findings on the importance of current account balance for the adoption of IT; further on, higher interest rates seem to increase the likelihood of introducing inflation rate though this finding need not be true for the long-term interest rates.

Concerning the exchange rate regime, there is a strong consensus that this monetary policy framework requires higher degree of exchange rate flexibility, i.e., currency pegs are not conducive to IT; in addition, there is a consensus that higher degree of central bank instrument-independence is a necessary condition for adopting IT; similarly, the empirical evidence generally supports the proposition that higher level of financial development is required for introducing IT though this finding does not receive uniform empirical support; finally, the empirical research on the importance of political institutions suggests that democracy, decentralization, and political polarization all increase the likelihood of adopting IT; on the other hand, the literature seems to be completely divided about the importance of trade and financial openness – some studies find that IT is more likely in more open economies, while others reach the opposite conclusion; similarly, the empirical evidence on the role of fiscal discipline is rather mixed, and this is equally true for the importance of both the budget balance and the public debt.

The lack of robust findings in this field is not surprising at all. In fact, both the early literature and the experience of inflation targeters suggest that the link between various macroeconomic and institutional determinants and the probability of adopting IT might be weak, i.e., they are not to be viewed either as strict necessary or sufficient conditions. For instance, many EMEs introduced the IT starting from moderate inflation rates, ranging from 10% to 40% (Mishkin and Schmidt-Hebbel 2002). Similarly, Brazil introduced the IT after the sharp devaluation in 1999, followed by fiscal and political instability (Mishkin 2004b, Mishkin and Savastano 2002). In the late 1990s, Poland, Hungary, and the Czech Republic adopted the IT notwithstanding the large fiscal deficits. In addition, during the initial phase, Poland and Hungary implemented the IT in the presence of exchange-rate bands and with a limited capacity for forecasting inflation (Jonas and Mishkin 2007). Therefore, the proponents of IT argue that the initial institutional and technical conditions as well as the macroeconomic environment are important but not critical for introducing IT, i.e., the lack of these conditions can be remedied after the introduction of this monetary regime (Batini and Laxton 2007, IMF 2006, Mishkin and Schmidt-Hebbel 2007). Given the lack of consensus in the empirical literature on the necessary preconditions for the implementation of IT, Neumann and von Hagen (2002) are probably right when concluding that the choice between IT and other monetary policy strategies is more a question of culture than economic considerations.

## 3. Macroeconomic effects of IT

Since the 2000s, there has been a growing interest in the effects of IT on macroeconomic performance, leading to diverse conclusions regarding the effectiveness of this monetary framework. This is particularly true for the empirical research focusing on EMEs, which represent a heterogeneous group with specific institutional and macroeconomic characteristics.

### 3.1. The effects on inflation expectations and inflation persistence

IT is a monetary regime in which price stability is the main objective of monetary policy, accompanied by explicit quantitative targets for the medium-term inflation rate. In addition, within this policy framework, the central bank is characterized by a high degree of transparency and accountability. It is believed that the announcement of explicit inflation targetc is instrumental in anchoring inflation expectations of financial markets participants and private agents in general. In this respect, if credible, the IT regime will affect the formation of inflation expectations by reducing the backward-looking component and making them more forward-looking. In turn, well-anchored inflation expectations will lead to a fundamental change in inflation dynamics by reducing or even eliminating inflation inertia.

Unsurprisingly, much of the empirical literature has been concerned with testing the presumed beneficial effects of IT on inflation expectations and inflation persistence. In this regard, most of the studies are based on surveys of inflation expectations of businesses, households, or professional forecasters, which are only available for OECD countries. In addition, some studies work with data extracted from bond markets. As for the methodological apparatus, apart from the OLS regression framework, a number of papers employ a variety of time-series analysis models, ranging from structural break tests and various autoregression models to GARCH and fractional integration. In what follows, we review the relevant literature on this subject matter.

Huh (1996) and Lane and Van Den Heuvel (1998) are examples of two early studies which examine the behavior of inflation expectations following the adoption of IT in the UK. Both papers cover similar time periods, spanning from mid-1970s to mid-1990s, employ similar econometric techniques – Vector Autoregression (VAR), and obtain the same finding – inflation expectations have declined following the adoption of IT. Within the VAR framework, Schmidt-Hebbel and Werner (2002) provide evidence on the favorable effects of IT on inflation expectations in Brazil, Chile, and Mexico, Demertzis et al. (2010) show that IT has contributed significantly to anchor inflation expectations in seven advanced economies, while Corbo et al. (2001) obtain the same results for a sample of 26 countries. Neumann and von Hagen (2002) conduct an event study of monetary policy for nine advanced economies by comparing the effects of the 1978 and 1998 oil shocks. They find that long-term interest rates, as a proxy for inflation expectations, increased in both IT and non-IT countries, but to a lesser amount in the former, concluding that inflation targeters achieved larger credibility gains. Johnson (2003), too, shows that the announcement of inflation targets has reduced inflation expectations in five advanced economies during 1984-1998. Gillitzer and Simon (2015) find that inflation expectations in Australia are better anchored in the latter phase of IT.

Gürkaynak et al. (2007) take an alternative approach to measuring the effects of IT on long-term inflation expectations, based on daily bond yields in Canada, Chile, and the USA. Specifically, they compare the behaviour of long-term nominal and indexed bond yields in response to important economic events and confirm the beneficial effects of IT. In a similar fashion, Gürkaynak et al. (2010) construct the zero-coupon yield curve separately for the nominal interest rates and for the real interest rates (for inflation-indexed bonds) in the USA, the UK, and Sweden. Then, they calculate the implied forward interest rates, and obtain the inflation compensation as a difference between nominal and real forward interest rates for a ten-year horizon. Finally, they regress inflation compensation on the surprise components in the announcements of important economic and financial data and conclude that, while US long-term inflation expectations are highly responsive to domestic economic news, the UK and Swedish ones generally respond neither to domestic nor to foreign news, implying that inflation expectations are firmly anchored in these two IT countries. De Pooter et al. (2014) follow the same methodology and find that inflation expectations are well-anchored in Brazil, Chile, and Mexico. Working with treasury yield data, Suh and Kim (2021) confirm the above findings for a mixed sample comprised of advanced countries and EMEs, providing some evidence that the anchoring effect might be even stronger in the latter.

Based on a sample of 15 countries, Ehrmann (2015) finds that inflation expectations in IT countries are better anchored compared with non-targeters. As a result, policy rates in IT countries need to react less to the changes in inflation so that IT countries are less likely to face the zero lower bound constraint. In addition, he studies the success of IT in anchoring inflation expectations under different circumstances: when inflation is normal, when it is persistently high, and when it is persistently low. He finds that when inflation is persistently low, the role of IT in anchoring inflation expectations get weaker, i.e., inflation expectations become more dependent on lagged inflation rate, the dispersion of individual inflation forecasts increases, and inflation expectations revise down when actual inflation is lower than expected, but they do not respond when actual inflation is higher than expected. Recently, Ehrmann (2021) investigates the performance of different types of inflation targets (targets set as ranges, points, and points surrounded by tolerance bands, respectively) and finds that no type of inflation target is superior in anchoring of inflation expectations.

However, the empirical literature is not unison about the favourable effects of IT on inflation expectations. Among the early papers, Freeman and Willis (1995) evaluate the effectiveness of IT in the early inflation targeters (the UK, Canada, New Zealand, and Sweden), and provide mixed evidence that inflation targets may have lowered inflation expectations in only two out of the four analysed countries. Similarly, Debelle (1996) finds that IT has led to lower inflation expectations in New Zealand, but the evidence for Canada is weak. Johnson (2002) analyses 11 developed countries and shows that, following the adoption of IT, inflation expectations declined immediately in New Zealand and Sweden, the effect is smaller and slower to develop in the case of Canada and Australia, while it vanishes in the United-Kingdom. Moreover, he finds that neither the variability of expected inflation nor the average absolute forecast error fell after the announcement of inflation targets. Levin et al. (2004) provide some evidence for the effectiveness of IT in anchoring long-run inflation expectations in industrialized countries, though it is not the case with short- and medium-run inflation expectations. Also, they employ the event-study approach to five EMEs and conclude that both short- and long-term inflation expectations in these countries did not change markedly after the introduction of IT. Working with the VAR methodology, Davis and Presno (2014) investigate inflation expectations for a sample of 36 developing countries and advanced economies. For the former, they find that inflation expectations respond less to shocks in inflation and oil prices during the post-IT period, thus, providing evidence that IT is beneficial in anchoring expectations. On the other hand, they obtain weak results for the latter sample. Similar results can be found in Capistrán and Ramos-Francia (2010), who examine the effect of IT on the dispersion of inflation expectations from professional forecasters for a panel of 25 countries. They find that, after controlling for the global inflation trend, disinflation periods, country-specific effects as well the level and variance of inflation, IT leads to lower long-term inflation expectations. However, the full favourable effect is visible only after three years following the adoption of IT; also, the anchoring effect is present only in the developing countries, but not in advanced economies.

A number of papers cast doubts about the effectiveness of IT in anchoring inflation expectations. For instance, on the basis of simple descriptive data for OECD countries, Almeida and Goodhart (1998) argue that the announcement of inflation targets did not affect inflation expectations, which continued to be either the same as before or higher than the announced inflation targets. Exploiting survey data on long-term inflation expectations for 15 advanced countries, Castelnuovo et al. (2003) find that inflation expectations are well anchored in almost all countries, notwithstanding whether they are inflation targeters or not, i.e., the announcements of explicit quantitative inflation targets does not matter for inflation expectations. This is indicated by both the low and generally decreasing volatility of expectations as well as the low and decreasing correlation between the revisions in short-term and long-term inflation expectations. Ball and Sheridan (2004) investigate the inflation performance of 20 OECD countries and find that inflation expectations behave similarly in both inflation targeters and non-targeters. Cecchetti and Hakkio (2010) and Willard (2012), also, show that IT does not reduce inflation expectations in OECD countries, Angeriz and Arestis (2007a) and Crowe (2010) provide similar evidence for mixed samples which include EMEs, while Filardo and Genberg (2009) obtain the same findings for inflation targeters in Asia. Further on, based on micro-level data for 10 090 firms from 81 countries, Broz and Plouffe (2010) find that exchange-rate pegs are effective in reducing private sector's concern with inflation expectations, while inflation targets and central bank independence are not. Recently, employing the event-study approach based on bond market data, Bundick and Smith (2018) compare the behavior of inflation expectations in the USA and Japan. They find that, in contrast to the US experience, forward measures of inflation compensation in Japan are responsive to news about current inflation after the introduction of IT, implying an absence of well-anchored inflation expectations.

The related strand of the empirical literature dealing with the effects of IT on inflation persistence provides equally conflicting evidence. Here, too, some papers show that the adoption of IT has caused a break in the behavior of the inflation process, other papers provide mixed evidence, while some of them clearly refute the beneficial effect of IT. For instance, within the first camp, Levin et al. (2004) provide evidence that inflation targeters in advanced economies exhibit lower inflation persistence in comparison with non-targeters. Kuttner and Posen (2001) compare the effects of hard pegs, central bank autonomy, and inflation targeting, and find that IT is the only monetary regime that reduces inflation persistence. Similarly, comparing the impact of IT and exchange rate targeting on inflation persistence for 68 countries, Kočenda and Varga (2018) find that both explicit and implicit IT are helpful in reducing inflation persistence. Baxa et al. (2014) and Bratsiotis et al. (2015) provide evidence for the advanced economies that inflation persistence is greatly reduced or even eliminated following the introduction of IT, while Pétursson (2004) and Phiri (2016) obtain the same results for 29 inflation targeters and 46 African countries, respectively. Finally, the beneficial effect of IT is confirmed in several papers focusing on individual countries, such as the UK (Kontonikas, 2004), Mexico (Chiquiar et al., 2010), South Africa (Gupta et al. 2017), and Turkey (Çiçek and Akar, 2013).

On the other hand, Kuttner and Posen (1999) find that inflation persistence in New Zealand and the UK declined after the adoption of IT, but IT had no effects whatsoever in Canada. Benati (2008) studies inflation persistence in five advanced economies under different monetary regimes and finds that, although inflation persistence declined after the adoption of IT, inflation exhibits very low persistence under the alternative nominal anchors, too, implying that IT cannot be regarded superior to other monetary regimes. Canarella and Miller (2016 and 2017a) study inflation persistence in OECD countries and shows that inflation persistence is lower after the adoption of IT in only half of the sample. In this regard, they find many asymmetries and variations in inflation persistence among the sample countries. Also, they provide mixed evidence on inflation persistence after the Global crisis, i.e., it increased in some countries, but not in others. In a subsequent paper, Canarella and Miller (2017b) show that inflation targeters among the industrial countries share a common persistence trend with Germany and the USA and, consequently, respond in the same way to inflationary shocks. That is, the industrial countries exhibit inflation dynamics moving in line with the global downward trend, but this is not true for the developing countries, where IT indeed leads to lower inflation persistence. Similarly, Yigit (2010) finds that inflation persistence declined in the post-IT period in only three out of eight advanced economies, while Davis and Presno (2014) show that IT is helpful in reducing inflation inertia in the developing countries, but not in the advanced economies. Working with a large sample of 91 countries, Vega and Winkelried (2005) provide some evidence that IT reduces inflation persistence in both advanced and developing countries, but the effect is both small and diminishes very quickly, i.e., it lasts for only one quarter. Gerlach and Tillmann (2012) show that IT reduces inflation persistence in Asia-Pacific, but not in the non-Asian inflation targeters. In their study of inflation dynamics in three CEE economies, Baxa et al. (2015) find that IT does not automatically change the dynamics of the inflation process, i.e., the effects of depend on the way IT is implemented, i.e., whether it is accompanied by targets for the exchange rate.

The findings in a number of papers challenge the proposition that IT may be an effective tool for reducing inflation persistence. For instance, Siklos (1999) and Levin and Piger (2004) show that IT does not lower inflation persistence in advanced economies, Ball and Sheridan (2004), Gadea and Mayoral (2006) and Willard (2012) obtain the same results for OECD countries, while Hossain (2014) confirms these findings for Australia. Edwards (2007), Siklos (2008), and Filardo and Genberg (2009) provide similar evidence working with mixed samples of advanced countries and EMEs. Similarly, Arsić et al. (2022) find that IT did not affect inflation persistence in a sample of 26 EMEs, Siregar and Goo (2010) confirms the ineffectiveness of IT in Indonesia and Thailand, while Kaseeram and Contogiannis (2011) show that IT has not been successful in reducing inflation inertia in South Africa.

Therefore, we can summarize that the empirical evidence in this field has failed to provide convincing evidence that IT itself serves as an effective tool for stabilizing inflation expectations and for reducing inflation persistence. In this regard, it is difficult to claim the superiority of IT over the alternative nominal anchors. In addition, even when IT does succeed in stabilizing inflation expectations, its impact appears with a considerable lag extending up to several years, which suggest that its favourable effects are largely dependent on the credibility of monetary policy.

### 3.2. The effects on inflation and inflation variability

Given that price stability is the predominant objective of central banks under IT, the empirical research on the effectiveness of this monetary regime has primarily focused on its impact on inflation performance (average inflation rate and inflation variability).

Due to the limited amount of data available at the time, some of the early papers in this field do not rely on formal econometric methodology, but draw their conclusions based on simple comparison between inflation targeters and non-targeters. For instance, Almeida and Goodhart (1998) compare inflation performance of 14 OECD countries and find that IT-countries do not have lower average inflation than non-targeters, while inflation variability is higher in IT-countries. On the other hand, comparing five inflation targeters with six non-targeters, Dotsey (2006) concludes that IT provides some benefits in reducing both inflation and its volatility.

At the same time, a number of papers attempt to identify the macroeconomic effects of IT based on the difference-in-difference methodology. Within this analytical framework, probably the most influential paper in the early empirical literature is Ball and Sheridan (2004). Working with a sample of 20 OECD countries, they regress various measures of economic performance (inflation, inflation variability, etc.) between the pre- and post-adoption periods on a dummy variable representing the treatment (the adoption of IT). However, they argue that this regression produces biased estimates due to the regression in mean, and, hence, they extend it by introducing another regressor – the average performance in the pre-targeting period as a proxy for the initial conditions. They show that IT did not lead to better inflation performance in IT countries relative to non-targeters, i.e., inflation targeters only converged to non-targeters which have already had lower inflation rates. In sum, Ball and Sheridan (2004) show that IT does not lead to better inflation performance, i.e., both inflation targeters and non-targeters have experienced improvement in economic performance during the analyzed period. Later on, employing the same methodology on an extended sample, Ball (2010) finds that IT leads to lower average inflation by 0.65 percentage points, which represents a non-negligible, but not a dramatic improvement. Similarly, Neumann and von Hagen (2002) study inflation performance of nine advanced countries and find that inflation targeters managed to reduce both the level and volatility of inflation in comparison with the pre-IT period, but they did not outperform non-targeters. Therefore, they are not able to prove the superiority of IT, concluding that inflation targeters only converged towards the monetary policy of non-targeting countries. On this ground, they claim that the choice between IT and other monetary policy strategies is more a question of culture than economic principles. On the other hand, Wu (2004) obtains opposite results, implying that IT leads to lower inflation in OECD countries.

Using several identification approaches, such as GMM, structural model, difference-in-difference, OLS, and IV, Willard (2012) confirms that IT does not affect inflation performance in OECD countries. De Mendonça (2007) estimates simple OLS cross-section regressions for 14 OECD countries. He compares the intercepts obtained from the pre- and post-IT regressions and, despite the statistical insignificance of the estimates, concludes that IT indeed provides lower inflation. Also, Parkin (2014) finds that CBI improves macroeconomic performance, but IT is a more effective arrangement for lowering inflation and inflation volatility.

Several papers investigate the effectiveness of IT in advanced economies by means of various time series analysis methods. For instance, employing three types of GARCH models, Kontonikas (2004) shows that the adoption of IT reduces long-run inflation uncertainty in the UK. Ftiti and Hichri (2014) analyze three IT-countries (UK, Sweden, and Canada) by the evolutionary spectral frequency methodology and find that IT leads to lower inflation. On the other hand, using ARMA, GARCH, and regime switching models, Genc et al. (2007) provide evidence that IT does not affect inflation in Canada, New Zealand, Sweden, and the UK. Based on the Kalman filter of the historical structural shocks consistent with an estimated DSGE model, Ravenna (2007) shows that IT did not contribute to the low inflation volatility in Canada since the early 1990s, concluding that low inflation volatility was the result of good luck, i.e., the low volatility of structural shocks. Similarly, Hossain (2014) finds that IT does not lower inflation volatility in Australia. Employing the VAR methodology and working with similar samples of advanced economies, both Laubach and Posen (1997) and Lee (1999) find that IT does not affect inflation. Similarly, Lane and Van Den Heuvel (1998) focus on the UK’s experience and show that inflation rate under the IT regime was not lower compared to the pre-IT period. Among the papers using Markov switching models, Choi et al. (2003) provide evidence that the IT regime may have had favourable effects on inflation volatility in New Zealand. On the other hand, Dueker and Fischer (2006) estimate a two-state

Markov switching model for six advanced countries. Their results suggest that inflation-targeting countries generally followed their non-inflation targeting neighbours in reducing trend inflation rates. Hence, they conclude that it is difficult to argue that formal inflation targets have led to any divergence between targeters and non-targeters in terms of inflation performance.

Finally, Lin and Ye (2007) investigate the macroeconomic effects of IT for 22 advanced economies by employing methodological approaches that are designed to address the self-selection problem. Specifically, they estimate several propensity score matching models and confirm the common finding that IT does not have any effects on inflation and inflation volatility. Ginindza and Maasoumi (2013) extend the traditional propensity score methodology and study the effects of IT on inflation and inflation volatility based on alternative matching models. Also, they estimate the entire distribution of potential outcomes instead of focusing only on the average treatment effect. Working with a panel of 30 advanced economies, they find that IT is associated with inflation rate, but it is difficult to quantify the precise effects. However, they show that late adopters of IT (those which implement this regime from five to ten years) do not utilize the beneficial effects on inflation, which are gained only by the early targeters (those with more than ten years of experience). Finally, they find small and insignificant effect of IT on inflation volatility.

As the popularity of IT has spread over the non-OECD, researchers were able to work with larger datasets consisting of both advanced and developing economies. Using various econometric techniques, such as IV models, pooled mean group estimator, and system-GMM, Calderón and Schmidt-Hebbel (2010) show that inflation targeters have lower inflation than non-targeters with the favourable effect being larger in developing countries. Kose et al. (2018) obtain the same results based on the difference-in-difference approach. Vega and Winkelried (2005) study 91 countries over 1990-2004 using the propensity score matching methodology. The estimated average treatment effects show that IT reduces average inflation and inflation variability in both advanced and developing countries, and for both soft and full-fledge targeters, with the effects being larger for developing countries and for soft-targeters. Two other papers based on propensity score matching models confirm these findings: Yamada (2013) finds that IT leads to lower inflation rate in comparison with flexible, intermediate, and fixed exchange rate regimes, while Mukherjee and Singer (2008) shows that IT is associated with lower average inflation in both OECD and non-OECD economies.

However, it seems that the empirical evidence based on mixed samples is far from convincing. For instance, working with a broad dataset of 124 monetary frameworks for 41 countries, Kuttner and Posen (2001) compare hard pegs, central bank autonomy, and IT, finding that, not only IT, but other monetary regimes, too, reduce inflation, suggesting that IT cannot be regarded a superior strategy. Hu (2006) finds that IT indeed leads to lower inflation, but the effect on inflation variability is not statistically significant. Similar results can be found in Abo-Zaid and Tuzemen (2012), who provide evidence on the beneficial effects of IT on average inflation rate in developing countries, while it does not matter in advanced economies; however, IT does not affect inflation volatility in both groups of countries. Samarina et al. (2014) and de Mendonça and de Guimarães e Souza (2012) confirm that the favourable effects on inflation are absent in the sub-sample of developed economies. Pétursson (2004) and Berument and Yuksel (2007) also find only limited evidence on the effectiveness of IT, which lowers inflation and inflation volatility only in part of the countries within their sample. Fratzscher et al. (2020) employ a dynamic panel-data model to investigate characteristics of IT as a shock absorber in times of large natural disasters. For the whole sample, they find that IT is associated with lower inflation, but it lowers inflation only in the sub-sample of OECD economies, while it does not have any effects in non-OECD countries. Ardakani et al. (2018) show that the results from the propensity score methodology may be quite sensitive to the specific estimation method. In these regards, when employing a nonparametric propensity score model, they find that IT actually leads to higher average inflation and inflation volatility in developing countries; however, they obtain opposite results from the parametric propensity score method, i.e. IT appears to reduce inflation volatility in these economies; finally, they employ a semi-parametric propensity score matching method and find that IT does not affect either inflation or inflation volatility in both the developed and developing economies.

In addition, several papers strongly reject the hypothesized effectiveness of IT as a tool for providing better inflation performance. Applying the intervention analysis to multivariate structural time series models, Angeriz and Arestis (2007a) find that IT does not affect inflation and inflation volatility. Arestis et al. (2014) find that inflation rates in OECD countries converge irrespective of their monetary policy frameworks, reflecting the global trend of declining inflation due to globalization. The estimates from the GMM and difference-in-difference models in Alpanda and Honig (2014) imply that IT does not affect inflation either in advanced or developing economies, while Filardo and Genberg, H. (2009) and Naqvi and Rizvo (2009) provide similar evidence for Asian inflation targeters.

A number of studies have focused exclusively on the effectiveness of IT in EMEs using a variety of econometric techniques. In these regards, Batini and Laxton (2007) argue that, in comparison with industrialized countries, the experience of emerging markets provide richer data set to analyze the effects of IT for several reasons: first, these countries offer shorter time span of the data, but this is compensated by the larger number of countries in both the treatment and control groups; second, EMEs usually are characterized by higher macroeconomic volatility during the pre-adoption period, which makes it easier to discern the effects of IT; third, these economies can provide more useful information about the performance of IT during periods of macroeconomic turbulence.

In their pioneering study, Batini and Laxton (2007) compare the experience of 13 inflation targeters with 29 countries with other monetary policy strategies. Employing the difference-in-difference methodology, they find that inflation targeters enjoy considerable gains in terms of both average inflation rate (by 4.8 percentage points) and inflation volatility (by 3.6 percentage points), implying that IT appears to be an effective strategy for controlling inflation in EMEs. The descriptive analysis in IMF (2006), Roger (2009), and Roger (2010) suggests that inflation targeters in EMEs have better macroeconomic performance in terms of average inflation and inflation variability. However, the findings from these papers should be taken with some caveats as they are based on a simple comparison of statistical data, i.e., they do not provide formal econometric evidence. Porter and Yao (2005) analyse the performance of IT "lite" in Mauritius, showing that it has led to a reduction in inflation rate. More recently, the beneficial effects of IT on inflation performance for larger samples of EMEs has been confirmed in several papers by means of different econometric techniques, such as Gemayel et al. (2011), Ouyang and Rajan (2016), and Arsić et al. (2022). Similarly, working with a large panel of 152 countries, Combes et al. (2014) investigate how the combination of IT and fiscal rules affect average inflation. In this regard, they show that, either on its own (irrespective of the existence of fiscal rules) or jointly with fiscal rules, IT is capable of delivering lower inflation. However, this favorable effect is present only when IT is adopted after the fiscal rules, but not the other way round.

However, the empirical literature has not provided consensus on the macroeconomics effect of inflation in EMEs, too. For instance, within the VAR methodology, Schmidt-Hebbel and Werner (2002) obtain mixed results concerning the effectiveness of IT in Brazil, Chile, and Mexico. Estimating several propensity score matching models for 52 developing countries, Lin and Ye (2009) find that IT reduces both the level of inflation and its volatility, though there is large heterogeneity in the effects of IT on inflation. Also, they show that the effects of IT depend on the country characteristics, such as the time length since the introduction of IT, the degree of fiscal discipline, the role of exchange rate regime etc. Employing the synthetic control methodology, Lee (2011), too, obtains mixed findings by showing that IT reduces inflation in only four of the ten inflation targeters. Further on, Ayres et al. (2014) find important regional differences in the effectiveness of IT in reducing inflation in EMEs, thus, failing to generalize its beneficial effects. In addition, the results from several empirical studies cast a serious doubt on the effectiveness of IT in reducing inflation volatility in EMEs. For instance, within the difference-in-difference methodology, Gonçalves and Salles (2008) show that the adoption of IT indeed leads to lower inflation in EMEs, but they fail to confirm the favorable effects on inflation volatility. The same findings are obtained by Brito and Bystedt (2010) and Stojanovikj and Petrevski (2021) by means of the GMM estimator. Baxa et al. (2015) study the effects of IT on inflation dynamics in three CEE economies using the Bayesian model averaging methodology. They find that the volatility of inflation declined sharply a few years after the adoption of IT in Czech and Poland, but it remained unchanged in Hungary. Based on the GARCH models, Kaseeram and Contogiannis (2011) conclude that IT has not been successful in reducing inflation volatility in South Africa. The estimates from the quantile regressions in Chevapatrakul and Paez-Farrell (2018) suggest that IT may have heterogeneous effects in EMEs. Specifically, they find that IT had negative effects on inflation only in the least successful countries, i.e., those that have managed to reduce inflation the least, while for the other quantiles IT has no effect on inflation whatsoever; also, they show that IT does not affect inflation volatility.

Further on, part of the empirical literature strongly rejects the hypothesis that IT improves inflation performance in EMEs. Revisiting the paper by Gonçalves and Salles (2008) and controlling for three additional elements, Thornton (2016) finds that IT does not reduce average inflation as compared to the countries with alternative monetary regimes. Recently, Duong (2021) obtains the same results employing the difference-in-difference methodology on a panel of 54 EMEs. Genc and Balcilar (2012) provide time series-based evidence that the adoption of IT in Turkey did not cause a structural break in inflation dynamics, i.e., actual inflation would not have been different from the inflation without IT. Finally, using several propensity score matching models for 16 CEE countries, Wang (2016) fails to provide evidence that IT is instrumental in reducing either inflation rate or inflation volatility.

Therefore, based on the results obtained from a variety of econometric techniques, we are able to conclude that the empirical literature has failed to provide convincing evidence on the beneficial effects of IT on inflation performance in advanced economies. According to Batini and Laxton (2007), the lack of strong evidence on the favorable effects of IT in these countries may reflect several factors, such as: the limited number of both inflation targeters and non-targeters, which results in a small data set for econometric inference; in addition, the macroeconomic performance of all advanced economies, inflation targeters and non-targeters alike, improved simultaneously during the 1990s for a variety of reasons; finally, most industrial countries have already achieved relatively low and stable inflation before the adoption of IT. The empirical evidence based on mixed samples is equally inconclusive though most of the papers suggest that the gains from the IT regime may have been more prevalent in the developing countries. This finding conforms with the meta-regression analysis of the existing literature on the macroeconomic effects of IT in Balima et al. (2020), who find that the papers working with samples consisting of developing countries are more likely to obtain significant negative effects of IT on inflation rate and its volatility. However, neither is this conclusion robust as demonstrated by the diverse results from the studies focusing exclusively on EMEs. The vast majority of this strand of empirical literature provide strong evidence against the hypothesis that the IT regime is effective in reducing inflation volatility, and even the presumed effects on reducing inflation are equally questionable. In the light of this conflicting evidence, it is safe to conclude that the question of whether inflation targeters have better inflation performance than countries that pursue alternative monetary strategies is still open.

### 3.3. The effects on output growth and volatility

In practice, IT is implemented in a flexible way in which the central banks do not behave like "inflation nutters". Instead, although price stability remains the predominant goal, the central banks often pursue some other short-run goals and seek to offset various shocks which occasionally hit the economy. As a result, the proponents of IT argue that it resembles optimal monetary policy in the sense of controlling inflation while simultaneously minimizing output fluctuations. Notwithstanding their theoretical attractiveness, as shown below, the empirical validity of these arguments remains an open issue.

In their pioneering study on the macroeconomic effects of IT in OECD countries, Ball and Sheridan (2004) show that this monetary regime does not matter either for average growth or output variability. In other words, they show that both inflation targeters and non-targeters have experienced improvement in economic performance during the analysed period. The same results are obtained by Ball (2010), who works essentially with same sample. Several other studies, employing different econometric methods, such as VAR (Laubach and Posen 1997, Mishkin and Posen 1997), cointegration (Lee 1999), and propensity scoring methods (Walsh 2009), also provide empirical evidence that IT does not affect output performance in advanced economies. Similarly, investigating the UK experience, Lane and Van Den Heuvel (1998) reach the same conclusion. Nadal-De Simone (2001) estimates the conditional variance of output in 12 advanced countries by employing a set of time-varying parameter state-space models. For Canada, he finds that output variance is higher in the post-IT period, for Australia and NZ he finds that IT leads to lower output volatility, while output variance does not change in Singapore and Korea.

On the other hand, several studies do provide evidence on the favourable effects of IT with respect to output performance in advanced economies. For instance, Choi et al. (2003) show that this monetary regime is associated with lower growth volatility in New Zealand. Based on a simple comparison of descriptive statistics for a sample of advanced economies, Dotsey (2006) concludes that IT provides some benefits in increasing output and lowering its volatility, while Parkin (2014) finds that IT is more effective than CBI in improving output growth and growth volatility. Recently, employing various panel-data models for a sample of 27 OECD countries, Ryczkowski and Ręklewski (2021) confirm the favourable effects of IT on output growth.

Even the empirical studies based on mixed samples, comprising both advanced countries and EMEs, fail to provide firm evidence on this issue. For instance, Corbo et al. (2001) show that, during 1980s and 1990s, inflation targeters were able to reduce the effects of inflation shocks on output volatility due to the anchored expectations. Also, Hu (2006) provides evidence that IT is associated with higher output growth and lower growth volatility, while Barnebeck Andersen et al. (2015) confirm the better performance of inflation targeters in terms of output growth. On the other hand, Pétursson (2004) obtains mixed results about the effects of IT on economic growth, whereas Naqvi and Rizvo (2009) find that IT does not affect either output growth or growth volatility.

The findings from a number of studies suggest that the output effects of IT may differ between advanced economies and EMEs. In this respect, the descriptive comparison between IT-countries and non-targeters in Roger (2010) implies that only low-income inflation targeters experienced improvement in growth, while among high-income countries, inflation targeters saw little change in macroeconomic performance. Similarly, Mollick et al. (2011) examine the effects of IT on output growth during the "globalization years" of 1986-2004. They find that, in industrial countries, both "soft" IT and full-fledged IT have positive effect on long-run growth, while in EMEs only full-fledge IT has positive effects on growth, probably reflecting the lower degree of credibility of "soft" IT. Abo-Zaid and Tuzemen (2012) find that IT raises growth rate in both developed and developing countries, but it lowers growth volatility only in the latter group. Fratzscher et al. (2020) find that IT improves output growth in the sub-sample of OECD economies, but not in non-OECD countries.

Batini and Laxton (2007) were the first to examine the macroeconomic effects of IT in EMEs. Concerning the effects on real output, they are not able to provide clear-cut results since they show that IT reduces output gap volatility, but the estimated effect is statistically insignificant. Nevertheless, they interpret this finding in a positive way by concluding that that IT does not have adverse effects on output in EMEs. Following the same methodology, Gonçalves and Salles (2008) support the proposition that IT leads to lower output volatility in these economies. Similarly, based on some descriptive statistics for a sample of 42 EMEs, IMF (2006) argues that inflation targeters have better macroeconomic performance in terms of output variability. Roger (2009) extends the comparisons in IMF (2006) and finds that, among low-income countries, inflation targeters experienced lower growth volatility than non-targeters during the 1990s and 2000s. However, the conclusion from these two studies should be taken with a caution since they do not provide formal econometric evidence.

Brito and Bystedt (2010) obtain some evidence that IT is not costless in EMEs, i.e., it has adverse effects on output growth whereas the effect on output volatility is insignificant; however, their evidence is rather weak as the estimated results vary considerably with the econometric methods employed. Amira et al. (2013), too, employ the GMM framework, but they provide opposite evidence by showing that IT indeed spurs economic growth in EMEs, though it is associated with higher growth volatility, too. The rest of the literature dealing with EMEs provides mixed evidence, too. For instance, employing different econometric methods, Gemayel et al. (2011) fail to provide robust findings on the effects of IT on GDP growth and growth volatility, while the growth effects of IT in Ayres et al. (2014) are statistically insignificant. Investigating the experience of developing countries, Ouyang and Rajan (2016) show that IT does not affect growth for the whole sample, but it leads to lower growth in the sample of Asian countries. Revisiting the paper by Gonçalves and Salles (2008) and controlling for three additional elements, Thornton (2016) fails to establish a significant association between IT and growth volatility. Chevapatrakul and Paez-Farrell (2018) obtain the same result using quantile regressions. Employing the difference-in-difference methodology, Duong (2021) finds that IT does not affect output growth. Arsić et al. (2022) also show that IT does not affect GDP growth, though it may be effective in reducing output volatility.

Therefore, the results from the empirical literature on the output effects of IT are very contradictory so that one can hardly summarize them. Obviously, the literature cannot provide convincing evidence that IT itself can enhance output growth. On the other hand, the available empirical evidence is not sufficient to claim that IT is harmful with respect to output performance either. Unsurprisingly, the ultimate conclusion depends on the way one interprets the above findings: the proponents of IT prefer to claim that it is able to maintain price stability without harming output, while the critics conclude that IT may not be viewed as a costless monetary strategy.

### 3.4. IT and the Global crisis

The Global financial and economic crisis of 2008 (Crisis) undermined the confidence of academic economists in IT as the most appropriate framework for conducting monetary policy (Reichlin and Baldwin, 2013). Commenting on the performance of IT during the Crisis, Walsh (2009) argues that IT does not constrain the central bank in counteracting the effects of the financial crisis; on the contrary, he claims that responding to financial turmoil is consistent with the goals of flexible IT, because a credible commitment to an inflation target prevents the danger of deflation. Also, he argues that flexible IT enables the central banks to cope with cost shocks (food and energy prices) and praises the decision of inflation targeters to breach inflation targets in the wake of the Crisis.

Notwithstanding the theoretical views on the ability of inflation targeters to cope with financial crises and cost shocks, this issue can be only resolved on empirical grounds. Recently, a number of papers have investigated the performance of IT during the Crisis, but they fail to provide firm conclusions on this issue.

For instance, de Carvalho Filho (2010) shows that, by lowering nominal and real interest rates and by experiencing sharp real depreciation, inflation targeters were less likely to face deflation though they did not outperform the non-targeters in terms of unemployment, industrial production, and output growth. Based on a large panel of more than 170 countries during 2007-2012, Rose (2014) compares the macroeconomic and financial effects of three monetary regimes: IT, exchange-rate pegs, and intermediary regimes. In this regard, he presents evidence that IT and exchange rate pegs produce similar outcomes with respect to several macroeconomic and financial variables, such as: the magnitude of business cycles, the size and volatility of capital flows, inflation, fiscal performance, output growth, current account, foreign reserves, real exchange rates, property prices, bond yields etc. Working with the same sample, Gagnon (2013), too, finds little difference between hard peggers and inflation targeters in terms of GDP growth. However, he shows that inflation targeters have better macroeconomic outcomes than hard peggers in terms of inflation variability. Fouejieu (2013) provides similar evidence for 79 countries, confirming that, in terms of inflation and GDP growth rate, there were no differences between IT countries and non-targeters during the Crisis; however, he finds that inflation targeters experienced lower inflation volatility and lower interest rates.

Comparing the performance of inflation targeters and non-targeters during the Crisis, Roger (2009, 2010) obtains different results for low-income and high-income countries: in the former, both inflation targeters and non-targeters saw similar decline in growth rates, but inflation rose to a lesser extent in IT countries; in the latter, inflation targeters saw smaller decline in growth rates and slightly less increase in inflation. Similarly, Fry-McKibbin and Wang (2014) provide mixed results by showing that IT have insulated developed countries from the global downturn, but the results are inconclusive for emerging markets in which IT has not proved to be successful regime during the Crisis.

In contrast, Barnebeck Andersen et al. (2015) find that, compared with other strategies, especially with fixed exchange rates, inflation targeters fared better during the Crisis in terms of output growth. The same conclusion is obtained by de Carvalho Filho (2011) for a sample of 51 advanced economies and EMEs. Specifically, he finds that inflation targeters have had better growth performance relative to non-targeters in the post-Crisis period; also, median inflation rate in inflation targeters was never lower than 1.5%, while it went below zero in non-targeters, suggesting that the former countries were less likely to face deflation. In addition, Kose et al. (2018) find that IT countries experienced higher real exchange rate volatility compared to non-targeters, while Kočenda and Varga (2018) show that inflation persistence in IT countries did not rise during the Crisis as it did in the non-targeters. Similarly, Duong (2021) finds that inflation targeters in EMEs had lower inflation than non-targeters during the Crisis. Finally, Arsić et al. (2022) show that the favourable effects of IT on inflation, inflation volatility, and GDP volatility were especially pronounced in the aftermath of the Crisis, but it did not affect either inflation persistence or GDP growth.

### 3.5. The effects on interest rates and exchange rates

Though most of the empirical literature deals with the macroeconomic effects of IT on inflation and output, few papers make a step forward and investigate the effects on other important policy variables, such as interest rates and exchange rates.

In principle, IT should be associated with lower nominal and real interest rates as well as with lower interest rate volatility provided that at least some of the following propositions are true: it has stabilizing effects on inflation expectations; it is effective in reducing average inflation rate and inflation volatility; it improves the inflation-output trade-off. However, the accumulated empirical evidence has not yet resulted in firm conclusions on this issue.

Among the early empirical papers, Huh (1996) and Lane and Van Den Heuvel (1998) analyse the UK experience within the VAR framework and find that both short- and long-term interest rates declined after the adoption of IT. Employing the same methodology, Laubach and Posen (1997), Mishkin and Posen (1997), and Neumann and von Hagen (2002) obtain the same results for a sample of several advanced economies. Similarly, Almeida and Goodhart (1998) estimate EGARCH models for six advanced economies and find that IT reduces short-term interest rate volatility.

On the other hand, several papers obtain opposite findings. For instance, Freeman and Willis (1995) evaluate the early experience with IT in UK, Canada, NZ, and Sweden. They find that IT has not have any effects on the time-series properties of nominal long-term interest rates, but it has led to higher long-term real interest rates. Chadha and Nolan (2001) estimate a state space model for the UK, showing that interest-rate volatility was higher during the second phase of IT, i.e., after the Bank of England had been granted the operational independence in May 1997. Interestingly, they interpret this rise in volatility as unavoidable consequence of the credible IT-regime, which leads to lower price-level volatility. Lee (1999) studies six advanced countries and finds that IT does not affect either short-term or long-term interest rates. Working with the difference-in-difference methodology, Wu (2004) finds that IT does not affect real interest rates in 22 OECD countries. In their well-known study, Ball and Sheridan (2004) show that there has been a downward trend in long-term interest rates in all OECD countries, both before and after the introduction of IT. This is also true for the volatility of short-term interest rates, which does not differ between inflation targeters and non-targeters. Similarly, Ball (2010) finds that IT actually increases both the level and variability of long-term interest rates. Finally, employing the propensity score matching methodology, Lin and Ye (2007) confirm that IT does not impact long-term nominal interest rates and their volatility in OECD countries, while Lee (2010) shows that short-term interest rates in inflation targeters are higher than those that would prevail had they followed either FED's or ECB's policy rules.

The findings from the empirical studies based on mixed samples are equally contradictory, though some of them support the hypothesis of ineffectiveness of IT in advanced economies. For instance, employing the propensity score matching methodology, Ardakani et al. (2018) find that the average treatment effect of IT on interest rate volatility is negative and statistically significant only in developing economies, but not in industrialized countries. Similarly, the estimates from the dynamic panel-data models in Fratzscher et al. (2020) imply that IT leads to higher policy rates in the sub-sample of OECD economies, but not in non-OECD countries. Based on the difference-in-difference methodology on a sample of ten Asian economies, Naqvi and Rizvo (2009) find that IT does not affect short-term interest rate volatility. On the other hand, Pétursson (2004) obtains opposite results for a sample of 29 inflation targeters, showing that IT reduces short-run nominal interest rates. Finally, IMF (2006) and Batini and Laxton (2007) argue that, in EMEs, inflation targeters seem to have experienced lower volatility in real interest rates. However, their conclusions should be taken cautiously since they are derived from simple descriptive statistics.

The relationship between IT and exchange rates can be analysed from two points of view: first, exchange rate regimes as a determinant in the decision process of whether to adopt IT; second, following the adoption of IT, its impact on the behaviour of nominal and real exchange rates. The first issue has been covered in Section 2 so that in what follows we focus on the latter research topic.

*A priori*, it is difficult to gauge the impact of IT on exchange rates: on the one hand, the implementation of IT requires flexible exchange-rate regime, which might result in excessive volatility of real exchange rates mostly driven by the changes in relative prices of tradable goods; on the other hand, the built-in flexibility of the IT regime, including more flexible exchange rates, may provide the central bank with more effective tools for curbing macroeconomic volatility; finally, under the assumption that IT indeed leads to lower inflation volatility, it will inevitably translate into lower real exchange rate variability due to the stability of internal prices. For instance, within a small open-economy model, Ball and Reyes (2008) show that credible IT regime is more similar to floating and managed floating regimes than to currency pegs or "fear of floating". That is, their model predicts that IT countries have lower probability of exchange rate changes than pure floaters, but higher probability than currency peggers, suggesting that, in terms of exchange rate variability, IT countries stand somewhere between floaters and peggers. On the other hand, Leitemo (2006) shows that in open economies IT may lead to higher nominal and real variability due to the long targeting horizons. Specifically, a long horizon of inflation forecast targeting results in extensive interest rate smoothing, which prevents nominal interest rates from responding to disequilibrium conditions. Consequently, interest rate smoothing results in higher inflation volatility, which translates to higher real interest rate and real exchange rate variability. In view of the above opposite arguments, this issue could be resolved only on empirical grounds.

Among the early evidence, Almeida and Goodhart (1998) estimate GARCH models for seven OECD countries over 1986-1997 and find that exchange-rate volatility during the IT regime has increased as compared to the fixed exchange-rate period, but it is lower than the period of discretionary monetary policy. Studying the UK's experience with IT based on a Bayesian VAR model, Lane and Van Den Heuvel (1998) find that IT has no effects whatsoever on exchange rates. Since exchange rate volatility may be more relevant for EMEs, the subsequent empirical research has focused on these countries, too. As a result, there are several studies that utilize mixed samples comprising both industrialized countries and EMEs. For instance, Rose (2007) studies whether inflation targeters suffer from higher exchange-rate volatility compared with non-targeters. Although the IT dummy is significantly negative in only 19 out of 66 variants of the regression model, nonetheless, he concludes that IT countries have lower exchange rate volatility than non-targeters. Employing propensity-score matching methods, Lin (2010) obtains similar results, i.e., he finds that IT reduces real exchange rate volatility for only one out of seven estimators, while it lowers nominal exchange rate variability for three estimators. The findings in Kose et al. (2018), too, cast doubts on the effectiveness of IT. Using the difference-in-difference approach, they find that IT has no impact on the level of real exchange rate; moreover, during 2007-2011, inflation targeters had higher volatility of real exchange rate than non-targeters, but then again, the effects of IT vanish during the 2011-2015 period. On the other hand, the estimates from the treatment regression in Pontines (2013) and the propensity score matching models in Ardakani et al. (2018) point to opposite conclusions, i.e., IT is associated with lower nominal and real exchange rate volatility.

Based on a broad dataset of 124 monetary frameworks for 41 countries during 1973-2000, Kuttner and Posen (2001) compare the effects of hard pegs, central bank autonomy, and IT on exchange rates. They find that both hard pegs and central bank autonomy reduce nominal exchange rate fluctuations, whereas the effects of IT are mixed. Similarly, Rose (2014) analyzes a large dataset of 170 countries over 2007-2012 and shows that both IT and currency pegs have more depreciated real exchange rates than intermediary regimes during the Crisis. Ouyang et al. (2016), employ propensity score matching, random-effects panel-data model and GMM to study the impact of IT, hard pegs and intermediate exchange rate regimes on real exchange rate volatility, relative tradables prices and tradables/nontradables prices for a panel of 62 developed and developing countries. For the whole sample, the estimates from all econometric methods imply that IT is associated with higher RER volatility as compared with both hard peggers and intermediate regimes whereas there are no differences between the monetary regimes in terms of external and internal RER volatility. Therefore, the findings in these papers suggest that IT cannot be regarded superior to other monetary frameworks.

Although the empirical research based on mixed samples has not been very helpful in obtaining firm conclusions on the effects of IT on exchange rate volatility, it suggests that IT may have differential effects on exchange rates in developed countries versus EMEs. For instance, Rose (2007) shows that while IT leads to higher exchange rate volatility in advanced countries, it is associated with lower variability in EMEs. The latter finding is consistent with the existence of dual objectives in EMEs where inflation targeters continue to smooth exchange rate fluctuations. The same results are obtained by Lin (2010), Ardakani et al. (2018) and Pontines (2013). Ouyang et al. (2016) confirm that in advanced economies IT is associated with greater real exchange rate volatility driven by external prices. As for developing countries, they find that IT does not affect real exchange rate volatility though they provide some evidence that it is associated with lower volatility in internal prices. Therefore, they hypothesize that in, developed countries, which pursue IT with floating nominal exchange rates, IT would be associated with higher real exchange rate volatility due to the higher nominal exchange rate volatility (driven by the relative prices of tradables). On the other hand, since developing IT countries usually operate under managed exchange rate regimes, they would be associated with lower real exchange rate volatility due to the stability in internal prices.

However, the above hypothesis about the beneficial effect of IT on exchange rate volatility has not found strong support in the empirical research focused strictly on EMEs. The early studies, such as IMF (2006) and Batini and Laxton (2007), provide some descriptive statistics implying that inflation targeters in EMEs seem to have experienced lower volatility in nominal exchange rates. On the other hand, based on GARCH models for several EMEs, Edwards (2007) finds that IT does not affect either nominal or real exchange rate variability. He offers a more IT-friendly interpretation of these findings: instead of concluding that IT does not lower exchange rate variability, he concludes that the adoption of IT regime per se does not increase the extent of exchange rate volatility in EMEs. Working with a panel of 37 EMEs, Berganza and Broto (2012) examine the effects of IT and foreign exchange intervention on real exchange rate volatility. While finding that exchange rates are more volatile under IT as compared to other monetary regimes, they show that foreign exchange interventions in some IT countries, mainly in Latin America, seem more effective in reducing exchange rate volatility than in non-targeting countries. Hence, they conclude that exchange rate volatility in the IT countries with managed exchange rate regimes tends to be lower than those with flexible exchange rates. Ouyang and Rajan (2016) investigate the impact of inflation targeting on real exchange rate volatility as well as in terms of its two component parts: relative tradable prices across countries (external prices) and sectoral prices of tradables and non-tradables within countries (internal prices). The main results from their study are the following: IT does not impact real exchange rate volatility; they provide some mixed evidence that IT lowers the volatility of real exchange rates for tradable goods; there is firm evidence that IT lowers internal real exchange rate volatility.

Therefore, the empirical literature in this field suggests that IT may be associated with higher exchange rate volatility in advanced economies, mainly due to the fact that it is accompanied by floating exchange rates. On the other hand, this monetary framework may have stabilizing effects on real exchange rate volatility in developing countries, which usually operate under managed exchange rate regimes. However, even this conclusion should not be taken at face value, because several studies suggest that IT cannot be regarded superior to hard pegs in containing excessive exchange rate volatility.

### 3.6. The effects on fiscal outcomes

As mentioned above, the early literature regards fiscal discipline one of the main prerequisites for adopting IT though the empirical support to this proposition is far from strong. However, this strand of empirical literature deals with the importance of fiscal variables on the likelihood to adopt IT, but not on the reverse relationship – how IT affects fiscal outcomes. In principle, the presumed favourable effects of IT on fiscal discipline may rationalized as follows (Minea and Tapsoba, 2014): first, fiscal authorities may have an incentive to improve on fiscal discipline in order to support the central bank's commitment to the inflation target; second, IT may improve fiscal performance by keeping inflation low, thus, mitigating the erosion of the real value of tax revenues (the negative Olivera-Tanzi effect); third, the lower inflation volatility associated with IT should stabilize the tax base, which in turn would result in better tax collection. However, note that the latter two arguments are conditional on the favourable effects of IT on average inflation and inflation volatility.

This issue is especially relevant for EMEs, which have a long tradition in relying on seignorage as a source of revenue. In this regard, Minea et al. (2021) build a theoretical model in which government spending is financed by both taxes and seignorage, and the quality of institutions affects tax collection. They show that, under some circumstances related to the political costs of policy reforms, IT may induce the government to offset the decline in seignorage by initiating political reforms (e.g., by improving the quality of institutions, fighting corruption etc.) in order to strengthen tax collection. Employing the propensity score matching method on a panel of 30 EMEs, Lucotte (2012) finds that the adoption of IT has quantitatively important effects on public revenue. Using similar methodology on a sample of 53 developing countries, Minea et al. (2021) confirms that IT may have favourable effects on tax collection.

Although the available empirical evidence is quite limited, generally, it supports the proposition that IT improves fiscal discipline. For instance, Miles (2007) investigates whether IT provides the same level of fiscal discipline as hard pegs. He finds that IT improves fiscal performance, but currency unions and currency boards (to a lesser extent) lead to tighter fiscal policy than inflation targets. Therefore, he concludes that IT does not provide better fiscal performance than exchange-rate pegs. Using the difference-in-difference methodology, Abo-Zaid and Tuzemen (2012) find that IT improves the budget balance in developed countries only, while it does not matter in developing countries; in addition, they find that IT does not affect budget balance volatility in both groups of countries. Combes et al. (2014) test several hypotheses related to the combination of IT and fiscal rules as well as their sequencing. Working with a panel of 152 countries over 1990-2009, they show that, taken on its own, i.e., irrespective of the existence of fiscal rules, IT has quantitatively important effects on the budget balance. When accompanied by fiscal rules, IT leads to even larger improvement in the budget balance, but this joint favorable effect is present only in the countries that have adopted fiscal rules before the adoption of IT.

Several papers deal with this topic within the propensity matching methodology. Minea and Tapsoba (2014) provide evidence that IT is associated with higher cyclically-adjusted fiscal balance and lower debt in developing countries, but not in developed ones. Fry-McKibbin and Wang (2014) find that inflation targeters in both developed economies and EMEs have lower tax burden and lower debt ratios than non-targeters, thus, supporting the favourable effects of IT on fiscal discipline. Ardakani et al. (2018), too, find that IT improves fiscal discipline in both developing and advanced economies with larger benefits in the latter. In the former, the adoption of IT encourages the fiscal authorities to pursue sound fiscal policy in order to support the credibility of the announced inflation targets. In the latter, governments are able to reap the credibility gains from the adoption of IT, which enables them to reduce the level of public debt. As a result, the favorable effect of IT on fiscal discipline is much larger in the advanced economies.

### 3.7. Methodological issues

Initially, the empirical literature on the effectiveness of IT relied on some form of an event-study analysis by comparing macroeconomic performance (inflation, output growth as well as their volatility) before and after the adoption of IT. For instance, in their pioneering study, Ball and Sheridan (2004) employ the difference-in-difference methodology, which estimates the effect of the treatment (IT) on the outcome (a macroeconomic variable such as the inflation rate) by comparing the change in the outcome variable in IT countries (between two periods: pre- and post-adoption dates) to the corresponding change in the same variable in the control group consisting of countries with other monetary regimes. However, this methodological approach may lead to misleading results if it does not take into account the change in macroeconomic variables that would occur anyway even in the absence of IT. This is the so-called "regression to the mean" problem. For instance, if there is a global downward trend in inflation during the period under study, this methodology would lead to biased results by inferring that IT has caused the decline in inflation in the targeting countries, while this outcome may have been driven by the global disinflation trend. To control for the "regression to the mean", Ball and Sheridan (2004) include the initial value of the outcome (e.g., the average macroeconomic performance in pre-adoption period) in the difference regression.

The estimates obtained from the difference-in-difference analysis may be sensitive to the definition of the pre- and post-intervention periods. Since various countries have adopted IT in different years, if one chooses a common initial period for a sample of inflation targeters, then the pre-intervention period would be different for each country. Also, the difference-in-difference methodology is characterized by a degree of arbitrariness with respect to determining the pre- and post-intervention periods for non-targeters. Although robustness checks and experimenting with a variety of initial periods may partly mitigate this issue, the obtained results continue to be sensitive to the choice of the initial period (Lee, 2011).

Another serious problem with the early empirical studies of macroeconomic effects of IT is their static nature as well as the failure to account for the potential endogeneity of IT. As for the former, the static models employed in the early papers may be largely misspecified due to the absence of dynamics. As for the latter criticism, the endogeneity of IT could arise from various sources, such as the effects of common shocks, omitted variables (unobserved heterogeneity), measurement errors, as well as reverse causality. For instance, Mishkin and Schmidt-Hebbel (2002) argue that the IT variable may be endogenous in the sense that the inflationary environment in EMEs may have led to the adoption of IT. As a result, the regression may not represent the causal effect of IT on macroeconomic performance, but it may only point to the existence of a mere statistical association between them.

In a response to these methodological issues, Brito and Bystedt (2010) have employed the General Method of Moments (GMM), which is a suitable technique for estimating dynamic panel data models. In these regards, dynamic panel-data models offer some advantages: first, they take a proper account of the unobserved heterogeneity within the sample and exploit the information offered by the large time dimension of the variables; second, working with panel-data models is a way to avoid the arbitrariness with respect to determining the initial period for non-targeters; third, by adding dynamics, the empirical model incorporates the entire history of the variables, so that regression coefficients, including the IT variable, represent the effect of new information. In addition, the GMM estimator has been designed to control for common time-variation, country-specific effects, and potential endogeneity of the variables in the regression model (see Arellano 2003, Bond 2002, Greene 2003, and Wooldridge 2002).

As already suggested, the main methodological problem in the estimation of true causal effects of IT on macroeconomic variables is that the assignment of the countries to the two groups (inflation targeters and non-targeters) is not random. On the contrary, conditional on some institutional and macroeconomic preconditions prevailing in a country, it decides whether to adopt the IT regime. As a result, the empirical studies are faced with the problem of selection bias, which arises when inflation targeters and non-targeters differ not only by the monetary policy strategy, but by other factors, too. In a response to this criticism, Lin and Ye (2007) have employed the propensity score matching methodology to estimate the macroeconomic effects of IT.

The propensity score matching is a quasi-experimental design which aims at estimating causal effects when the units are not assigned randomly to treatment. It consists of matching inflation targeters with non-targeters on the basis of some observed characteristics which are not affected by the treatment (the choice of IT). Then, the causal effect of IT on macroeconomic performance is calculated as the mean difference in the outcome (e.g., inflation, output growth etc.) across the treated and control group (inflation targeters and non-targeters, respectively). The analysis within this framework involves two stages: first, estimating the propensity score, i.e., the conditional probability of adopting IT controlling for a number of macroeconomic and institutional variables which seem to be relevant for this choice; second, estimating the average treatment effect on the treated by matching the treatment group (inflation targeters) with the control group (non-targeters) based on the propensity scores. In this way, the propensity score matching methodology allows for balancing the differences between inflation targeters and non-targeters. Here, it is understood that the binary model in the first stage (probit or logit) is properly specified as otherwise the propensity scores would be estimated poorly.

Though propensity score matching offers some advantages compared to the alternative econometric methods, it provides valid estimates only if two assumptions are satisfied: the conditional independence, and the common support condition (Khandker et al. 2010). The former requires that the binary model in the first stage must include all the variables that affect both the assignment to treatment (the choice of IT) and the potential outcomes. This condition implies that the choice of IT is based only on the countries' observable characteristics and not on some unobservable factors (such as society's preferences). The latter assumption implies that each inflation targeter can be matched to a counterfactual from the non-targeting countries based on the propensity score, which requires that the propensity score distribution is similar in the two groups of countries. Lee (2011) argues that the application of propensity score matching might be challenging when working with panel-data sets consisting of aggregate entities, such as countries, where propensity scores are estimated in two ways: either using a set of covariates for each country that are measured prior to the adoption of IT or using one-period lagged covariates relative to each country-year observation. Unfortunately, both approaches are not flawless: the former option implies that a variable measured at some point in time would have time-invariant effects thereafter, while the latter option might compromise the conditional independence assumption by having matching variables affected by the treatment. Consequently, both approaches introduce a bias in the measurement of the average treatment of the treated. In addition, Ginindza and Maasoumi (2013) criticize the traditional propensity score matching as being arbitrary in terms of the score matching in the sense that using the same multiple indicators can produce different scores. Hence, instead of estimating the average treatment effect, they present the whole distribution of potential outcomes, thus, emphasizing the uncertainty in the quantification of treatment effects.

In an attempt to circumvent the above-mentioned problems, Lee (2011) employs the synthetic control group approach, developed by approach of Abadie and Gardeazabal (2003) and Abadie et al. (2010). This methodology is similar to propensity score matching in that both approaches rely on the construction of a control group. However, within the synthetic control group methodology, based on the observable covariates during the pre-treatment period, for each individual inflation targeter a synthetic matching country is constructed which represents a good counterfactual for the treated country in the pre-intervention period. Then, the effects of IT are estimated as the difference between the outcomes between the inflation-targeting country and the synthetic group during the post-intervention period. By using the observable pre-treatment characteristics instead of country-year observations as the unit of analysis, the synthetic control group method is not prone to the potential problems that can bias the matching estimates. According to Lee (2011), the synthetic control method is more appropriate for analyzing the effects of IT for the following reasons: first, it provides country-specific estimates of the effects of IT, rather than an average treatment effect; second, given the country-by-country estimation of treatment effects, the definition of the post-intervention period for the control group is straightforward; third, it addresses explicitly the issue of selection on idiosyncratic temporary shocks in the pre-intervention period.

In their comprehensive meta-regression analysis, Balima et al. (2020) show that the macroeconomic effects of IT are quite sensitive to the estimation techniques employed. Specifically, they find that the papers employing propensity score matching, instrumental variables, and GMM tend to obtain significant results more often than the papers employing difference-in-difference methods. In these regards, the former three methods find more often that IT leads to lower inflation and inflation volatility, while the papers employing the latter methodology more often find opposite results. On the other hand, GMM and difference-in-difference tend to obtain similar results about the effects of IT on growth volatility, while propensity score matching more often find that IT increases growth volatility. Concerning the growth effect of IT, they show that it does not matter whether one uses GMM or propensity score matching, while the studies using difference-in-difference report less significant results.

### 3.8. A note on the interpretation of empirical findings

Methodological concerns aside, when reviewing the empirical literature on macroeconomic effects of IT, a brief comment on the substantial degree of subjectivity in the interpretation of empirical evidence is worthwhile. In these regards, it is obvious that, notwithstanding the actual estimates obtained from econometric models, some researchers seem to draw conclusions that confirm with their prior beliefs. In other words, the proponents of IT tend to emphasize the beneficial effects of IT even when they are not supported by firm econometric evidence. We provide several examples to illustrate this practice.

For instance, Corbo et al. (2001) argue that the adoption of IT has led to a change in the monetary policy reaction function as a result of the greater credibility. Specifically, they estimate Taylor-rule equations and find that both the coefficient of inflation and the coefficient of output gap declined in IT countries during the 1990s, while there is no change in non-targeters. Hence, they conclude that inflation targeters have managed to reduce the response of interest rates to inflation and output gap even though both regression coefficients for the group of IT countries are not statistically significant. In another example, Rose (2007) studies whether inflation targeters suffer from higher exchange-rate volatility compared with non-targeters. In this regard, he estimates 66 variations of the regression model finding that IT reduces exchange rate volatility in only 19 specifications. Although it is obvious that this evidence can be regarded weak or, at most, mixed, he concludes that inflation targeters experience lower exchange rate volatility than non-targeters. Further on, he argues that IT countries face less sudden-stops episodes compared to both the pre-IT period and to non-targeting countries although the estimated coefficients of the differences are statistically insignificant. Similarly, Edwards (2007) investigates the effects of IT on exchange rate volatility in seven countries. He finds that IT reduces nominal exchange-rate variability in three out of seven countries, notwithstanding that in all cases the regression estimates are close to zero. Nonetheless, he concludes that the adoption of IT regime per se does not increase the extent of exchange rate volatility.

Another example of subjective interpretation of econometric evidence is Gupta et al. (2010), who, ignoring the lack of statistical significance of impulse responses from the VAR model as well as the very low percent (almost zero) of inflation forecast errors explained by monetary policy shocks, argue that the effectiveness of monetary policy in South Africa has increased after the adoption of IT. In a similar fashion, one of the main conclusions in Abo-Zaid and Tuzemen (2012) is that fiscal imbalances are significantly improved when countries, especially developed ones, target the inflation rate, even though the regression coefficient is statistically insignificant. Finally, assessing the impact of IT on inflation and output performance in CEE countries, Wang (2016) reports that IT has favourable effects on GDP per capita although the estimated average treatment effects are significant in only two out of four matching methods.

The meta-regression analysis of the existing literature on the macroeconomic effects of IT in Balima et al. (2020) points to one source of this subjectivity in empirical research. Specifically, they find two sources of publication bias: first, authors and publishers are more likely to publish studies with statistically significant estimates about the effects of IT on inflation, output growth and their volatilities; second, the studies are more likely to be published if they show favourable effects of IT on inflation volatility and output growth though no such bias exists about the effects of IT on average inflation and on growth volatility.

## 4. IT and Disinflation Costs

As mentioned above, the proponents of IT emphasize its favorable effects on inflation expectations via, at least, three channels: first, the firm commitment to the announced inflation target accompanied by central bank independence improves the credibility of monetary policy makers; second, it is characterized by close communication with the public and higher degree of transparency, in general; third, within this strategy, central bank responds to the deviations from the inflation target only gradually in order to minimize output fluctuations. Consequently, by anchoring inflation expectations and by pursuing flexible policy, IT is expected to result in lower disinflation costs (Agénor, 2002).

The empirical literature in this field has been pioneered by Ball (1994), who proposed a simple measure of the foregone output due to disinflation policy (the so-called sacrifice ratio). His approach is based on the following assumptions about the behaviour of trend output in OECD countries: output is at its trend level at the start of a disinflation episode; output is at its trend level four quarters after an inflation trough; and output grows log-linearly between these two points when trend and actual output are equal. In addition, his approach relies on trend inflation to identify peaks and troughs, i.e., the period from a particular peak to the trough is identified as a disinflation episode. Then, the sacrifice ratio is calculated by dividing the cumulated output loss (the sum of deviations of actual output from trend output) by the change in trend inflation during each disinflation episode. Ball (1994) estimated the sacrifice ratios for a sample of 19 OECD countries during 1960-1991, finding that its main determinants are the speed of disinflation and nominal wage rigidity.

This method, although far from being flawless, has been widely used in the empirical research. In these regards, two additional variations of the original approach proposed by Ball (1994) have emerged in the subsequent literature. Criticizing the Ball's measure on the ground that it does not include the long-lasting effects accompanying each disinflation episode, Zhang (2005) proposes a slightly modified measure of the sacrifice ratio, relying upon the Hodrick-Prescott filter, and assuming that potential output grows throughout the particular episode at the rate implied by the Hodrick-Prescott filter at the beginning of the disinflationary episode. Further on, Hofstetter (2008) builds upon this approach with the additional assumption that output is at its trend level one year prior to the year the disinflation episode starts, thus trying to capture an even larger portion of the longer-lasting effects of disinflation.

In addition to the Ball's procedure and its variants, there are other possible approaches to measure disinflation costs. For instance, one approach is based on the estimates of the Phillips curve, whose slope measures the inflation-output trade-off, i.e., the response of output to the changes in inflation (Andersen and Wascher 1999, Gordon et al. 1982, Gordon 2013, Hutchison and Walsh 1998). An alternative method employs a structural VAR model to estimate the effects of monetary policy on output and inflation (based on the impulse responses of output and inflation to a monetary policy shock) and then the sacrifice ratio is calculated as a ratio between the cumulative impulse responses of output and inflation over some specified horizon (Cecchetti and Rich 2001, Belke and Böing, 2014).

Over time, a large body of empirical literature has emerged dealing with the determinants of sacrifice ratios in both advanced and developing countries. Some of the papers investigate the determinants of sacrifice ratios in general (Hofstetter 2008, Katayama et al. 2019, and Senda and Smith 2008), while others focus on various specific factors, such as central bank independence (Baltensperger and Kugler 2000, Brumm and Krashevski 2003, Daniels et al. 2005, Diana and Sidiropoulos 2004, Jordan 1997), trade openness (Bowdler 2009, Daniels and VanHoose 2006, 2009, 2013, Temple 2002), central bank transparency (Chortareas et al. 2003), political factors (Caporale 2011, Caporale and Caporale 2008), labour market institutions (Bowdler and Nunziata 2010, Daniels et al. 2006), and fiscal variables (Durham 2001). However, within this strand of empirical literature, there is much less evidence on the relationship between IT and the sacrifice ratios, and this is especially true for EMEs. In what follows, we provide a brief overview of the main findings from these studies, while the details are given in Table 4 of the Appendix.

The empirical literature based on the experience of OECD countries provides a unanimous conclusion that IT does not lead to lower sacrifice ratios. For instance, Debelle (1996) calculates the sacrifice ratios in New Zealand, Canada, and Australia, and finds that, in all cases, sacrifice ratios in the post-1989 period are higher than before. Laubach and Posen (1997) study eight advanced countries and find that in all IT-countries sacrifice ratios were larger than the average from previous disinflation episodes. In contrast, the most recent sacrifice ratios in non-targeting countries were lower than those during previous disinflation episodes. Similarly, in their study of disinflation cost in 13 OECD countries, Almeida and Goodhart (1998) find that both IT-countries and non-targeters had higher sacrifice ratios in the 1990s than in the 1980s. For the 1990s, the simple comparison between IT-countries and non-targeters (without providing formal econometric evidence) shows that the former countries have higher sacrifice ratio measured in unemployment terms. In terms of output, the sacrifice ratio is lower in IT-countries than in non-targeters, but this is true for both 1980s and 1990s. However, if UK is excluded from the sample, then the sacrifice ratios do not differ across IT-countries and non-targeters. Based on the estimates from the short-run Phillips curves for 21 OECD countries, Chortareas et al. (2003), too, provide evidence that IT does not affect sacrifice ratios. Also, the often-cited study of Ball and Sheridan (2004) show that this monetary regime appears to have had no significant effect on the disinflation process in 20 OECD countries. Finally, Roux and Hofstetter (2014) show that IT reduces sacrifice ratios only if disinflation is slow (more than three years), while it is irrelevant in fast disinflations. It seems that Gonçalves and Carvalho (2009) is the only study suggesting that adopting IT in OECD countries makes disinflation policy less costly. However, Brito (2010) criticizes their methodological approach that compares disinflations in inflation-targeting countries with non-simultaneous disinflations in non-targeting countries which occurred under different macroeconomic conditions. Therefore, controlling for the common trends in economic conditions, he obtains opposite results, i.e., disinflation costs under the IT regime are even larger compared to non-targeters.

The empirical literature that utilizes mixed samples comprising advanced and developing countries fails to provide firm evidence on the effectiveness of IT in reducing disinflation costs. For instance, Cecchetti and Ehrmann (2002) find that the average cross-country sacrifice ratio of inflation targeters is larger than the average sacrifice ratio in non-targeting countries though within both groups of countries the estimated sacrifice ratios vary considerably. Following the same approach and a similar sample, Corbo et al. (2002) show that IT countries had similar GDP-based sacrifice ratios before and after the adoption of IT. Tunali (2008) studies how various features of the IT regime affect the sacrifice ratio and obtain mixed results: the aggregate IT index lowers the sacrifice ratios only for the whole sample, but not for the two sub-samples; in addition, he finds that none of the specific features of IT (policy focus, accountability, instrument independence, financing government deficits, the use of forecasting and simulation methods etc.) are not statistically significant. Working with a large sample of 189 countries, Mazumder (2014) finds that the effects of IT are statistically insignificant in both OECD and developing countries. On methodological grounds, the study in Ardakani et al. (2018) differs sharply from the rest of the empirical literature by estimating the average treatment effects of IT based on the propensity score methodology. However, they fail to provide convincing evidence on the potential benefits of IT as a disinflation device, i.e., their results appear to be very sensitive to the specific estimation technique employed: the estimates form the semiparametric propensity score matching method imply that IT lowers the sacrifice ratios in developed economies only, and not in EMEs; yet, when employing the nonparametric series propensity score model, they find that IT reduces the sacrifice ratio in the full sample, but not in the individual sub-samples; finally, in the parametric propensity score method the effects of IT are insignificant in the full sample as well as in the two sub-samples. Magkonis and Zekente (2020) estimate the sacrifice ratios for a panel of 42 countries employing the Bayesian model averaging methodology. They find IT does not affect the sacrifice ratios, and this is true for both OECD and non-OECD countries. On the other hand, Gonçalves and Carvalho (2008) show that IT is generally associated with lower sacrifice ratios with the beneficial effects being larger in OECD countries. Recently, Sethi and Acharya (2019) study whether IT enhanced central bank credibility in 13 Asian countries and find that IT is associated with lower sacrifice ratios.

The empirical research of sacrifice ratios focusing exclusively on EMEs is very scarce. For instance, employing the Philips curve-based approach for a sample of EMEs, Brito and Bystedt (2010) provide evidence that the effects of IT are not statistically significant. On the other hand, based on a cross-country study of 44 EMEs, Stojanovikj and Petrevski (2020) provide evidence that adopting IT is associated with higher sacrifice ratios.

Therefore, although the main features of IT promise to offer a less costly device for controlling inflation, the empirical support to this proposition seems to be rather weak. Also, note that the empirical research on the determinants of sacrifice ratios, including the effects of IT, is burdened with a number of methodological issues. For instance, it is known that the estimates of sacrifice ratios are sensitive to the specific method applied in the estimation of potential output. In these regards, a number of papers demonstrate that the magnitude of sacrifice ratios differs considerably when trend output is estimated by alternative methods (see Temple 2002, Senda and Smith 2008, and Mazumder 2014). Another, though minor, issue is which period should be considered a disinflation episode. Most of the empirical papers follow Ball (1994) in identifying as disinflation episodes only the periods in which trend inflation has fallen by at least 1.5 or 2 percentage points from peak to trough, but this approach is clearly arbitrary. In addition, working with annual data, particularly for the inflation rate, provides a challenge in adopting the original approach of Ball (1994). His estimates of sacrifice ratios are based on both quarterly and annual data for OECD countries, but the former are not available for many EMEs. Obviously, data availability poses an obstacle in the precise identification of disinflation episodes. Further on, the original approach of Ball (1994) employs a method for estimating sacrifice ratio that discards supply shocks which, if not relevant for advanced economies, remain to be important determinants of sacrifice ratios in EMEs. It is well-known that these countries are frequently exposed to adverse external shocks, which result in large inflation-output volatility (Blanchard and Simon 2001, Fraga et al. 2003, Kose et al. 2003, Ramey and Ramey, 1995). And last, but not least, the estimation techniques that have been employed in most of the empirical literature do not tackle properly some important issues, such as endogeneity and selection bias. Therefore, more appropriate methods are required for inferring the true causal effects of IT on disinflation costs.

## 5. Conclusions

This paper surveys the abundant empirical literature related to the implementation of IT regime. Specifically, the paper focuses on three main issues: the main institutional, macroeconomic, and technical determinants that affect the adoption of IT (especially in EMEs); the effects of IT on macroeconomic performance (inflation expectations, inflation persistence, average inflation rate, inflation variability, output growth, output volatility, interest rates, exchange rates, and fiscal outcomes); and disinflation costs of IT (the so-called sacrifice ratio).

The existing literature points to a number of economic, institutional and technical prerequisites for adopting IT, such as: the absence of fiscal dominance, strong external position, relatively low inflation, developed financial markets and sound financial system, central bank independence, some structural characteristics (price deregulation, low dollarization, low sensitivity to supply shocks, strong external position etc.), the absence of de facto exchange rate targets, well developed technical infrastructure for forecasting inflation etc. While theoretically sound, the empirical research has not led to firm conclusions on the importance of each individual determinant for the adoption of IT, reflecting the fact that developed countries and EMEs represent heterogeneous groups with different institutional and macroeconomic characteristics.

However, most of the empirical studies suggest that larger countries and those with higher level of economic development are more likely to adopt this monetary regime; these findings imply that IT may not be a feasible monetary regime for small and/or low-income countries. Also, there is strong empirical evidence that the countries experiencing higher past inflation are less likely to switch to IT; this finding is consistent with the common requirement that the introduction of IT is conditional on previous disinflation. As for the exchange-rate regimes, there is strong consensus that IT requires higher degree of exchange rate flexibility, i.e., currency pegs are not conducive to IT. In addition, the empirical research generally support the proposition that central bank instrument-independence and higher level of financial development are necessary conditions for adopting IT. Finally, the empirical research on the importance of political institutions suggests that democracy, decentralization, and political polarization all increase the likelihood of adopting IT. On the other hand, the empirical literature seems to be completely divided about the importance of trade and financial openness, the role of fiscal discipline (budget balance and public debt). Similarly, the available research has provided diverse results on the effects of economic activity, i.e., it is not clear whether the countries facing more (less) stable economic conditions (measured by output growth and output volatility) are good (bad) candidates to implement IT.

The lack of robust findings in this field is not surprising at all. In fact, the literature suggests that the link between various macroeconomic and institutional determinants and the likelihood of adopting IT may be rather weak, i.e., they are not to be viewed either as strict necessary or sufficient conditions. In addition, the actual experience of inflation targeters, especially in the EMEs, confirms that a country need not meet all these preconditions, at least in the initial phase. More importantly, it seems that the adoption of IT itself promotes the fulfilment of these preconditions. In this respect, there is some empirical support to the hypothesis of a structural change after the adoption of IT, implying that, even when a country does not meet all the preconditions, once it has adopted it, this decision leads to changes in the institutions which support the functioning of IT.

Much of the empirical literature has been concerned with testing the presumed beneficial effects of IT on inflation expectations and inflation persistence. In this regard, most of the studies are based on surveys of inflation expectations with a few of them working with data extracted from bond markets. The existing empirical evidence in this field has failed to provide convincing evidence that IT itself may serve as an effective tool for stabilizing inflation expectations and for reducing inflation persistence. In addition, even when IT does succeed in stabilizing inflation expectations, its impact appears with a considerable lag extending up to several years, which suggest that its favourable effects are largely dependent on the credibility of monetary policy.

Given that price stability is the predominant objective of central banks under IT, the empirical research on the effectiveness of this monetary regime has primarily focused on its impact on inflation performance (average inflation rate and inflation variability). The research focused on advanced economies has failed to provide convincing evidence on the beneficial effects of IT, concluding that inflation targeters only converged towards the monetary policy of non-targeters. The lack of strong evidence on the favorable effects of IT in these countries may reflect several factors, such as: the limited number of both inflation targeters and non-targeters, which results in a small data set for econometric inference; in addition, the macroeconomic performance of all advanced economies, inflation targeters and non-targeters alike, improved simultaneously during the 1990s for a variety of reasons; finally, most industrial countries have already achieved relatively low and stable inflation before the adoption of IT. The empirical evidence from the mixed samples is equally inconclusive though most of the papers suggest that the gains from the IT regime may have been more prevalent in the developing countries. In the light of the conflicting empirical evidence, it is safe to conclude that the question of whether inflation targeters have better inflation performance than countries that pursue alternative monetary strategies is still open.

In practice, IT has been implemented in a flexible way meaning that, although price stability remains the predominant goal, central banks often pursue some other short-run goals and seek to offset various shocks which occasionally hit the economy. As a result, the proponents of IT argue that it resembles optimal monetary policy in the sense of controlling inflation while simultaneously minimizing output fluctuations. Notwithstanding their theoretical attractiveness, the empirical validity of these arguments remains an open issue. Specifically, the results from the empirical research on the output effects of IT are very contradictory so that one can hardly summarize them. Obviously, the literature cannot provide convincing evidence that IT itself can enhance output growth. On the other hand, the available empirical evidence is not sufficient to claim that IT is harmful with respect to output performance either. Unsurprisingly, the ultimate conclusion depends on the way one interprets the above findings: the proponents of IT prefer to claim that it is able to maintain price stability without harming output, while the critics conclude that IT may not be viewed as a costless monetary strategy.

Though most of the empirical literature focuses on the macroeconomic effects of IT on inflation and output, few papers make a step forward and investigate the effects on other policy variables, such as interest rates and exchange rates. In principle, IT should be associated with lower nominal and real interest rates as well as with lower interest rate volatility provided that at least some of the following propositions are true: it has stabilizing effects on inflation expectations; it is effective in reducing average inflation rate and inflation volatility; it improves the inflation-output trade-off. However, the accumulated empirical evidence has not yet resulted in firm conclusions on this issue. Similarly, a priori, it is difficult to gauge the impact of IT on exchange rates: on the one hand, the implementation of IT requires flexible exchange-rate regime, which might result in excessive volatility of real exchange rates mostly driven by the changes in relative prices of tradable goods; on the other hand, the built-in flexibility of the IT regime, including more flexible exchange rates, may provide the central bank with more effective tools for curbing macroeconomic volatility; finally, under the assumption that IT indeed leads to lower inflation volatility, it will inevitably translate into lower real exchange rate variability due to the stability of internal prices. Although the empirical research has not been very helpful in obtaining firm conclusions on the effects of IT on exchange rate volatility, it suggests that IT may have differential effects on exchange rates in advanced economies versus EMEs: on the one hand, IT may be associated with higher exchange rate volatility in advanced economies, mainly due to the fact that it is accompanied by floating exchange rates; on the other hand, this monetary framework may have stabilizing effects on real exchange rate volatility in developing countries, which usually operate under managed exchange rate regimes. However, even this conclusion should not be taken at face value, because several studies suggest that IT cannot be regarded superior to hard pegs in containing excessive exchange rate volatility.

Recently, the empirical literature has dealt with the effects of IT on the behaviour of fiscal authorities. In principle, the presumed favourable effects of IT on fiscal discipline may rationalized as follows: first, fiscal authorities may have an incentive to improve on fiscal discipline in order to support the central bank's commitment to the inflation target; second, IT may improve fiscal performance by keeping inflation low, thus, mitigating the erosion of the real value of tax revenues; third, the lower inflation volatility associated with IT should stabilize the tax base, which in turn would result in better tax collection. However, note that the latter two arguments are conditional on the favourable effects of IT on average inflation and inflation volatility. Although the empirical evidence on the effects of IT on fiscal policy is quite limited, generally, it supports the idea that IT indeed improves fiscal discipline.

In principle, by anchoring inflation expectations and by pursuing flexible policy, IT is expected to result in lower disinflation costs. The empirical research based on the experience of inflation targeters in advanced economies provides a unanimous conclusion that IT does not lead to lower sacrifice ratios. On the other hand, the empirical literature that utilizes mixed samples comprising advanced and developing countries fails to provide firm evidence on the effectiveness of IT in reducing disinflation costs. Therefore, although the main features of IT promise to offer a less costly device for reducing inflation, the empirical support to this proposition seems to be rather weak. In addition, the empirical research in this field is burdened with a number of unresolved issues such as the uncertain estimates of potential output, precise identification of disinflation episodes, the incorporation of supply shocks in the estimation of sacrifice ratios. Therefore, more appropriate methods are required for inferring the true causal effects of IT on disinflation costs.

Therefore, despite the vast empirical evidence on macroeconomic effects of IT, the findings from these studies are rather inconclusive: while some papers suggest that IT is associated with better macroeconomic performance (well-anchored inflation expectations, lower average inflation, improved inflation/output variability, fiscal discipline, etc.), others show that it does not produce superior macroeconomic benefits or, at most, they are quite modest. That said, it seems that the increasing popularity of IT is not based on strong empirical evidence with respect to the macroeconomic performance of this regime.

A general conclusion from the survey of empirical literature is that the macroeconomic effects of IT are quite sensitive to the estimation techniques employed. Further on, methodological concerns aside, there is a substantial degree of subjectivity in the interpretation of the empirical evidence. In these regards, it is obvious that, notwithstanding the actual estimates obtained from econometric models, some researchers seem to draw conclusions that confirm with their prior beliefs. In other words, the proponents of IT tend to emphasize the beneficial effects of IT even when they are not supported by firm econometric evidence. One important source of this subjectivity in empirical research may be the publication bias, i.e., authors and publishers are more likely to publish studies with statistically significant estimates about the effects of IT on inflation. Finally, much of the empirical research on macroeconomic effects of IT is burdened by some important methodological issues related to their failure to account for the potential endogeneity of this monetary regime. Specifically, the IT variable in regression models may be endogenous in the sense that conditional on some institutional and macroeconomic preconditions prevailing in a country, it decides whether to adopt the IT regime. As a result, the regression results may not represent the causal effect of IT on macroeconomic performance, but it may only point to the existence of a mere statistical association between them. In a response to this criticism, recent studies have employed estimation techniques designed to deal with the problem of selection bias, such as propensity score matching methodology or synthetic control group approach.

Appendix

Table 1. The adoption of IT in selected industrialized countries and EMEs

| Country | Year of adoption |
|---|---|
| | Industrialized countries |
| Australia | 1993 (Ardakani et al. 2018, Combes et al. 2014, Fraga et al 2003, Gerlach and Tiemmann 2012, Kose et al. 2018, Levin et al. 2004, Lin 2010, Pétursson 2004, Rose 2007, Samarina and De Haan 2014, Thornton and Vasilakis 2017)<br>1994 (Ball and Sheridan 2004, Canarella and Miller 2017a, Combes et al. 2014, Corbo et al. 2001, Corbo et al. 2002, Mishkin and Schmidt-Hebbel 2002, Samarina and De Haan 2014, Vega and Winkelried 2005) |
| Canada | 1990 (Gerlach and Tiemmann 2012)<br>1991 (most of the literature)<br>1992 (Combes et al. 2014)<br>1994 (Samarina and De Haan 2014, Vega and Winkelried 2005)<br>2001 (Gerlach and Tiemmann 2012) |
| New Zealand | 1989 (Ardakani et al. 2018)<br>1990 (Batini and Laxton 2007, Canarella and Miller 2017a, Combes et al. 2014, Corbo et al. 2002, Freedman and Laxton 2009, Gerlach and Tiemmann 2012, Kose et al. 2018, Leyva 2008, Lin 2010, Mishkin and Schmidt-Hebbel 2002, Pétursson 2004, Rose 2007, Samarina and De Haan 2014)<br>1991 (Samarina and De Haan 2014, Vega and Winkelried 2005)<br>2001 (Thornton and Vasilakis 2017) |
| Sweden | 1992 (Gerlach and Tiemmann 2012)<br>1993 (Ardakani et al. 2018, Canarella and Miller 2016, Canarella and Miller 2017a, Combes et al. 2014, Corbo et al. 2002, Kose et al. 2018, Lin 2010, Pétursson 2004, Rose 2007)<br>1995 (Combes et al. 2014, Thornton and Vasilakis 2017, Vega and Winkelried 2005) |
| UK | 1992 (Ardakani et al. 2018, Canarella and Miller 2017a, Combes et al. 2014, Corbo et al. 2002, Gerlach and Tiemmann 2012, Kose et al. 2018, Lin 2010, Pétursson 2004, Rose 2007, Vega and Winkelried 2005)<br>1993 (Ball and Sheridan 2004, Leyva 2008) |
| | 2000 (Thornton and Vasilakis 2017) |
| Finland | 1993 (Combes et al. 2014, Corbo et al. 2002, Freedman and Laxton 2009, Lin 2010, Mishkin and Schmidt-Hebbel 2002, Rose 2007, Vega and Winkelried 2005)<br>1994 (Combes et al. 2014) |
| Spain | 1994 (Mishkin and Schmidt-Hebbel 2002, Samarina and De Haan 2014)<br>1995 (Combes et al. 2014, Corbo et al. 2002, Freedman and Laxton 2009, Lin 2010, Rose 2007, Samarina and De Haan 2014) |
| | EMEs |
| Chile | 1990 (Pétursson 2004)<br>1991 (Canarella and Miller 2016, Canarella and Miller 2017a, Combes et al. 2014, Corbo et al. 2002, Gerlach and Tiemmann 2012, Gonçalves and Salles 2008, Lin 2010, Levin et al. 2004, Lucotte 2012, Mishkin and Schmidt-Hebbel 2002, Rose 2007, Samarina and De Haan 2014)<br>1999 (Ardakani et al. 2018, Batini and Laxton 2007, Combes et al. 2014, Freedman and Laxton 2009, Kose et al. 2018, Lucotte 2012, Samarina and De Haan 2014, Thornton and Vasilakis 2017, Vega and Winkelried, 2005) |
| Colombia | 1995 (Samarina and De Haan 2014)<br>1999 (most of the literature)<br>2000 (Gonçalves and Salles 2008, Lucotte 2012, Samarina and De Haan 2014,) |
| Czech Republic | 1997 (Ardakani et al. 2018, Kose et al. 2018)<br>1998 (most of the literature) |
| Ghana | 1992 (Thornton and Vasilakis 2017)<br>2002 (Ardakani et al. 2018)<br>2007 (Combes et al. 2014, Freedman and Laxton 2009, Kose et al. 2018, Samarina and De Haan 2014) |
| Indonesia | 2000 (Gerlach and Tiemmann 2012)<br>2005 (Ardakani et al. 2018, Combes et al. 2014, Kose et al. 2018, Leyva 2008, Samarina and De Haan 2014, Rose 2007, Thornton and Vasilakis 2017) |
| Israel | 1992 (Ardakani et al. 2018, Canarella and Miller 2016, Canarella and Miller 2017a, Combes et al. 2014, Corbo et al. 2002, Gonçalves and Salles 2008, Levin et al. 2004, Leyva 2008, Lin 2010, Lucotte 2012, Pétursson 2004, Rose |
| | 2007, Mishkin and Schmidt-Hebbel 2002, Samarina and De Haan 2014)<br>1997 (Batini and Laxton 2007, Combes et al. 2014, Freedman and Laxton 2009, Kose et al. 2018, Leyva 2008, Lucotte 2012, Samarina and De Haan 2014, Vega and Winkelried, 2005)<br>2001 (Gerlach and Tiemmann 2012, Thornton and Vasilakis 2017) |
| Mexico | 1995 (Samarina and De Haan 2014, Thornton and Vasilakis 2017)<br>1999 (Canarella and Miller 2016, Canarella and Miller 2017a, Combes et al. 2014, Corbo et al. 2002, Gonçalves and Salles 2008, Levin et al 2004, Lin 2010, Mishkin and Schmidt-Hebbel 2002, Pétursson 2004, Rose 2007, Vega and Winkelried, 2005)<br>2001 (Ardakani et al. 2018, Combes et al. 2014, Freedman and Laxton 2009, Kose et al. 2018, Lucotte 2012, Samarina and De Haan 2014)<br>2002 (Batini and Laxton 2007) |
| Peru | 1989 (Thornton and Vasilakis 2017)<br>1994 (Corbo et al. 2002, Gonçalves and Salles 2008, Mishkin and Schmidt-Hebbel 2002, Leyva 2008, Samarina and De Haan 2014)<br>2002 (Ardakani et al. 2018, Batini and Laxton 2007, Combes et al. 2014, Levin et al. 2004, Lin 2010, Pétursson 2004, Rose 2007, Samarina and De Haan 2014, Vega and Winkelried 2005) |
| Philippines | 1995 (Samarina and De Haan 2014)<br>2002 (most of the literature) |
| Poland | 1998 (Ardakani et al. 2018, Combes et al. 2014, Corbo et al. 2002, Freedman and Laxton 2009, Kose et al. 2018, Levin et al. 2004, Lin 2010, Pétursson 2004, Rose 2007)<br>1999 (Batini and Laxton 2007, Gonçalves and Salles 2008, Leyva 2008, Samarina and De Haan 2014) |
| South Africa | 2000 (Ardakani et al. 2018, Canarella and Miller 2016, Canarella and Miller 2017a, Combes et al. 2014, Corbo et al. 2002, Gerlach and Tillmann 2012, Gonçalves and Salles 2008, Levin et al. 2004, Leyva 2008, Lin 2010, Pétursson 2004, Rose 2007, Samarina and De Haan 2014)<br>2002 (Thornton and Vasilakis 2017)<br>2005 (Combes et al. 2014, Leyva 2008) |
| South Korea | 1997 (Thornton and Vasilakis 2017)<br>1998 (Ardakani et al. 2018, Canarella and Miller 2016, Canarella and Miller 2017a, Combes et al. 2014, Corbo et al. 2002, Gonçalves and Salles 2008, Levin et al. 2004, Leyva 2008, Lin 2010, Pétursson 2004, Samarina and De Haan 2014)<br>1999 (Gerlach and Tillmann 2012)<br>2001 (Freedman and Laxton 2009, Kose et al. 2018, Samarina and De Haan 2014) |
| Thailand | 2000 (most of the literature)<br>2006 (Thornton and Vasilakis 2017) |
| Turkey | 2000 (Thornton and Vasilakis 2017)<br>2002 (Samarina and De Haan 2014)<br>2006 (most of the literature) |

Author's compilation from various sources: Ardakani et al. (2018), Batini and Laxton (2007), Ball and Sheridan (2004), Bernanke and Mishkin (1997), Bernanke et al. (1999), Canarella and Miller (2016), Canarella and Miller (2017a), Combes et al. 2014, Corbo et al. (2002), de Mendonca and de Guimarães e Souza (2012), Freedman and Laxton (2009), Gerlach and Tillmann (2012), Gonçalves and Salles (2008), Kose et al. (2018), Levin et al. (2004), Leyva (2008), Lin (2010), Lucotte (2012), Mishkin and Schmidt-Hebbel (2002), Pétursson (2004), Rose (2007), Samarina and De Haan (2014), Thornton and Vasilakis (2017), and Vega and Winkelried (2005).

Table 2A. Selected empirical studies on the determinants of adopting IT

| Study | Data and methodology | Findings |
|---|---|---|
| Ardakani et al. (2018) | 98 advanced and developing countries during 1998-2013; binary response model | GDP growth (-) in all samples,<br>inflation (-) in all samples,<br>money growth (-) in all samples,<br>fixed exchange rate (-) in the whole sample and in developing countries,<br>central bank's assets (+) in the whole sample and in advanced countries, and (-) in developing countries,<br>financial development (-) in all samples. |
| Arsić et al. (2022) | 26 EMEs during 1997–2019; probit model | inflation (-),<br>current account balance (-),<br>fixed exchange rate regime (-),<br>trade openness (-),<br>money growth (-),<br>population size (+),<br>public debt (+). |
| Carare and Stone (2006) | Advanced countries and EMEs; cross-section regression and ordered probit model | GDP per capita (+) in the whole sample, and (x) in EMEs,<br>inflation (x),<br>CBI (x) in the whole sample, and (-) in EMEs,<br>financial development (+),<br>public debt (x) in the whole sample, and (-) in EMEs,<br>fiscal balance (x). |
| de Mendonça and de Guimarães e Souza (2012) | 180 advanced and developing countries during 1990-2007; probit model | GDP per capita (+),<br>monetization (+),<br>financial openness (+),<br>fiscal balance (-),<br>inflation (-),<br>fixed exchange rate (-),<br>trade openness (x). |
| Fouejieu (2017) | 26 EMEs during 2000-2010; probit model | inflation (-),<br>real GDP (-),<br>GDP growth (x),<br>fixed exchange rate (-),<br>trade openness (x),<br>long-term interest rate (x),<br>short-term interest rate (+),<br>CBI (+). |
| Fry-McKibbin and Wang (2014) | 31 advanced countries and 60 EMEs during 2007-2012; logit model | inflation (+) in advanced countries, and (-) in EMEs,<br>fixed exchange rates (-) in advanced countries,<br>trade openness (x) in advanced countries,<br>openness (-) in EMEs,<br>fixed exchange rates (-) in EMEs,<br>money growth (x) in both samples. |
| Hu (2006) | 66 advanced and developing countries during 1980-2000; logit model | output growth (-),<br>inflation (-),<br>real interest rates (+),<br>external debt (-),<br>fiscal balance (+),<br>floating exchange rate regime (x),<br>CBI (+),<br>pressure on exchange rates (+),<br>growth variability (x),<br>output gap (x),<br>nominal int rate (x),<br>nominal and real exchange variability (x),<br>current account (x),<br>terms of trade (x),<br>trade openness (x),<br>financial depth (x),<br>central bank autonomy (x). |
| Ismailov et al. (2016) | 82 advanced and developing countries in 2010; probit model and multivariate logit model | floating exchange rate (+) in the whole sample, (+) in advanced countries, and (x) in developing countries,<br>public debt (-) in the whole sample, (x) in advanced countries, and (-) in developing countries,<br>inflation (x) in the whole sample, (+) in advanced countries, and (x) in developing countries,<br>political risk (x) in all samples. |
| Leyva (2008) | 28 advanced countries and EMEs during 1975-2005; logit and probit models | inflation (-),<br>financial development (+),<br>GDP per capita (+),<br>trade openness (+),<br>fiscal balance (x). |
| Lin (2010) | 74 advanced countries and EMEs during 1985-2005; probit model | fiscal balance (+),<br>fixed exchange-rate regime (-),<br>inflation (-),<br>money growth (-),<br>per capita GDP growth (x),<br>trade openness (-) in developing countries. |
| Lin and Ye (2007) | 22 advanced countries during 1985-1999; probit model | inflation rate (-),<br>money growth (-),<br>CBI (+),<br>fixed exchange rate regime (-),<br>fiscal balance (x),<br>per capita GDP growth (x),<br>trade openness (x). |
| Lin and Ye (2009) | 52 developing countries during 1985-2005; probit model | inflation (-)<br>money growth (-)<br>fixed exchange rate regime (-)<br>trade openness (-)<br>real GDP per capita growth (x)<br>debt/GDP ratio (x)<br>CBI (x). |
| Lucotte (2010) | 30 EMEs during 1986-2005; probit model | CBI (+),<br>inflation (-),<br>government and overall political stability (+),<br>political polarization (+),<br>number of veto players (+),<br>decentralization (+),<br>GDP per capita (+),<br>exchange rate flexibility (+),<br>trade openness (+),<br>financial development (x). |
| Lucotte (2012) | 30 EMEs during 1980-2004; probit model | GDP growth (+),<br>flexible exchange rate (+),<br>CBI (+),<br>inflation (-),<br>trade openness (x),<br>financial development (x),<br>number of inflation targeters (+). |
| Minea et al. (2021) | 53 developing countries during 1984-2007; probit model | inflation (-),<br>tax revenue (-),<br>trade openness (-),<br>GDP per capita (+),<br>exchange rate flexibility (+),<br>constraints on the executive (+),<br>institutional quality (-), |
| | | primary schooling (+). |
| Minea and Tapsoba (2014) | 84 advanced and developing countries during 1985-2007; probit model | inflation (-),<br>trade openness (-)<br>CBI (+),<br>debt/GDP ratio (-),<br>fixed exchange rate (-),<br>fiscal rules (+). |
| Mishkin and Schmidt-Hebbel (2002) | 27 advanced and developing countries during 1990s; probit model | inflation (+),<br>trade openness (+),<br>instrument CBI (+),<br>goal-CBI (-),<br>legal CBI (x),<br>money targets (-),<br>fiscal balance (x). |
| Mukherjee and Singer (2008) | 49 OECD and non-OECD countries during 1987-2003; probit model | GDP growth variability (+),<br>floating exchange rate (+),<br>real interest rate (+),<br>inflation (+),<br>current account balance (+),<br>right-wing government (+),<br>central bank without bank regulatory authority (+). |
| Pontines (2013) | 74 advanced and developing countries during 1985-2005; treatment effects regression | inflation (-),<br>fiscal surplus (-),<br>money growth (-),<br>exchange rate flexibility (+),<br>GDP growth (x). |
| Rose (2014) | 170 advanced and developing countries during 2007-2012; multinomial logit model | country size (+),<br>democracy (+),<br>trade openness (x),<br>financial openness (x). |
| Samarina and De Haan (2014) | 60 OECD and non-OECD countries during 1985-2008; probit model and multinominal probit model | inflation (-) in the whole sample and in OECD countries, and (x) in non-OECD countries,<br>GDP growth (-) in the whole sample and in OECD countries, and (x) in non-OECD countries,<br>output volatility (+) in the whole sample and in OECD countries, and (x) in non-OECD countries,<br>flexible exchange rate (+),<br>exchange rate volatility (+),<br>CBI (x) in the whole sample and in OECD countries, and (+) in non-OECD countries,<br>fiscal balance (x), |
| | | public debt (-) in the whole sample and in OECD countries,<br>external debt (x),<br>financial development (-) in the whole sample and in OECD countries,<br>trade openness (x),<br>financial stability (x),<br>financial structure (x). |
| Samarina and Sturm (2014) | 60 advanced and developing countries during 1985-2008; random-effects probit model | inflation (-)<br>output volatility (-)<br>flexible exchange rate (+)<br>exchange rate volatility (+)<br>government debt (x)<br>financial development (+). |
| Samarina et al. (2014) | 85 advanced and developing countries during 1985-2011; logit model | advanced countries:<br>GDP per capita (-) in advanced countries, and (+) in developing countries,<br>financial development (-) flexible exchange rates (+) in both samples,<br>trade openness (+) in advanced countries, and (-) in developing countries,<br>inflation (-) in both samples,<br>financial openness (x) in advanced countries, and (+) in developing countries,<br>public debt (x) in advanced countries, and (-) in developing countries,<br>money growth (-) in developing countries. |
| Stojanovikj and Petrevski (2019) | 44 EMEs during 1990-2017; logit model | inflation (x),<br>inflation volatility (-),<br>growth (x),<br>GDP growth volatility (-),<br>financial development (+),<br>CBI (+),<br>capital mobility (+),<br>public debt (-). |
| Thornton and Vasilakis (2017) | 90 advanced and developing countries during 1979-2014; probit model | inflation (-) in all samples,<br>GDP per capita growth (+) in the whole sample and in developing |
| | | countries, and (x) in advanced countries,<br>public debt (-) in all samples,<br>trade openness (+) in the whole sample, (-) in advanced and in developing countries,<br>financial openness (+) in the whole and in developing countries, and (x) in advanced countries,<br>exchange rate flexibility (+) in all samples,<br>financial development (+) in all samples,<br>financial crises (-) in developing countries. |
| Vega and Winkelried (2005) | 91 advanced and developing countries during 1990-2004, logit model | investment/GDP (+)<br>fiscal balance (x)<br>inflation (+)<br>inflation volatility (-)<br>monetization (+)<br>trade openness (-)<br>floating exchange regime (-). |
| Wang (2016) | 16 EMEs during 1990-2010, logit model | population size (+),<br>inflation (x)<br>GDP growth (x)<br>trade openness (x)<br>debt/GDP ratio(x)<br>IT-neighbor (x)<br>money growth (x). |
| Yamada (2013) | 121 EMEs and developing countries during 1995-2007; multinomial logit model; | GDP level (+),<br>output gap (x),<br>time trend (+),<br>country size (+),<br>changes in terms of trade (+),<br>money growth (-),<br>trade openness (x),<br>foreign reserve (x). |

Note: “+” and “-“ indicate higher and lower likelihood to adopt IT, respectively, while “x” indicated that the factor is statistically insignificant or economically negligible.

Table 2B. Summary of the main findings on the determinants of adopting inflation targeting

| Determinant | Findings | Sample |
|---|---|---|
| GDP / GDP per capita | Carare and Stone (2006) (+) | Mixed sample |
| | Carare and Stone (2006) (x) | EMEs |
| | de Mendonça and de Guimarães e Souza (2012) (+) | Mixed sample |
| | Fouejieu (2017) (-) | EMEs |
| | Hu (2006) (x) | Mixed sample |
| | Ismailov et al. (2016) (x) | Mixed sample |
| | Leyva (2008) (+) | Mixed sample |
| | Lucotte (2010) (+) | EMEs |
| | Minea et al. (2021) (+) | Developing countries |
| | Samarina et al. (2014) (-) | Advanced countries |
| | Samarina et al. (2014) (+) | Developing countries |
| | Yamada (2013) (+) | Developing countries |
| GDP growth | Ardakani et al. (2018) (-) | Mixed sample |
| | Ardakani et al. (2018) (-) | Advanced countries |
| | Ardakani et al. (2018) (-) | Developing countries |
| | Fouejieu (2017) (x) | EMEs |
| | Hu (2006) (-) | Mixed sample |
| | Lin (2010) (x) | Mixed sample |
| | Lin and Ye (2007) (x) | Advanced countries |
| | Lin and Ye (2009) (x) | Developing countries |
| | Lucotte (2012) (+) | EMEs |
| | Pontines (2013) (x) | Mixed sample |
| | Samarina and De Haan (2014) (x) | Mixed sample |
| | Thornton and Vasilakis (2017) (+) | Mixed sample |
| | Thornton and Vasilakis (2017) (x) | Advanced countries |
| | Thornton and Vasilakis (2017) (+) | Developing countries |
| | Wang (2016) CEE (x) | EMEs |
| GDP variability | Hu (2006) (x) | Mixed sample |
| | Mukherjee and Singer (2008) (+) | Mixed sample |
| | Samarina and De Haan (2014) (+) | Mixed sample |
| | Samarina and De Haan (2014) (+) | Advanced countries |
| | Samarina and De Haan (2014) (x) | Developing countries |
| | Samarina and Sturm (2014) (-) | Mixed sample |
| | Stojanovikj and Petrevski (2019) (-) | EMEs |
| Inflation | Ardakani et al. (2018) (-) | Mixed sample |
| | Arsić et al (2022) (-) | EMEs |
| | Fry-McKibbin and Wang (2014) (+) | Advanced countries |
| | Fry-McKibbin and Wang (2014) (-) | EMEs |
| | Hu (2006) (-) | Mixed sample |
| | Lin (2010) (-) | Mixed sample |
| | Lin and Ye (2007) (-) | Advanced countries |
| | Lin and Ye (2009) (-) | Developing countries |
| | Minea et al. (2021) (-) | Developing countries |
| | Minea and Tapsoba (2014) (-)<br>Pontines (2013) (-)<br>Samarina and De Haan (2014) (-)<br>Samarina and De Haan (2014) (-)<br>Samarina and De Haan (2014) (x)<br>Samarina and Sturm (2014) (-)<br>Samarina et al. (2014) (x)<br>Samarina et al. (2014) (-)<br>Thornton and Vasilakis (2017) (-)<br>Vega and Winkelried (2005) (+)<br>Wang (2016) (x) | Mixed sample<br>Mixed sample<br>Mixed sample<br>Advanced countries<br>Developing countries<br>Mixed sample<br>Advanced countries<br>Developing countries<br>Mixed sample<br>Mixed sample<br>EMEs |
| Money growth | Ardakani et al. (2018) (-)<br>Ardakani et al. (2018) (-)<br>Ardakani et al. (2018) (-)<br>Arsić et al (2022) (-)<br>Fry-McKibbin and Wang (2014) (x)<br>Lin (2010) (-)<br>Lin and Ye (2007) (-)<br>Lin and Ye (2009) (-)<br>Pontines (2013) (-)<br>Samarina et al. (2014) (-)<br>Wang (2016) (x)<br>Yamada (2013) (-) | Mixed sample<br>Advanced countries<br>Developing countries<br>EMEs<br>Mixed sample<br>Mixed sample<br>Advanced countries<br>Developing countries<br>Mixed sample<br>Developing countries<br>EMEs<br>Developing countries |
| Real interest rates | Hu (2006) (+)<br>Mukherjee and Singer (2008) (+) | Mixed sample<br>Mixed sample |
| Long-term interest rates | Fouejieu (2017) (x) | EMEs |
| Short-term interest rates | Fouejieu (2017) (+)<br>Hu (2006) (x) | EMEs<br>Mixed sample |
| Fixed exchange rates | Ardakani et al. (2018) (-)<br>Ardakani et al. (2018) (x)<br>Ardakani et al. (2018) (-)<br>Arsić et al (2022) (-)<br>de Mendonça and de Guimarães e Souza (2012) (-)<br>Fouejieu (2017) (-)<br>Fry-McKibbin and Wang (2014) (-)<br>Lin (2010) (-)<br>Lin and Ye (2007) (-)<br>Lin and Ye (2009) (-)<br>Minea and Tapsoba (2014) (-) | Mixed sample<br>Advanced countries<br>Developing countries<br>EMEs<br>Mixed sample<br>EMEs<br>Mixed sample<br>Mixed sample<br>Advanced countries<br>Developing countries<br>Mixed sample |
| Exchange rate flexibility / Floating exchange rates | Hu (2006) (x)<br>Ismailov et al. (2016) (+)<br>Ismailov et al. (2016) (+)<br>Ismailov et al. (2016) (x)<br>Lucotte (2010) (+)<br>Lucotte (2012) (+) | Mixed sample<br>Mixed sample<br>High-income countries<br>Low-income countries<br>EMEs<br>EMEs |
| | Minea et al. (2021) (+)<br>Mukherjee and Singer (2008) (+)<br>Pontines (2013) (+)<br>Samarina and De Haan (2014) (+)<br>Samarina and Sturm (2014) (-)<br>Samarina et al. (2014) (+)<br>Samarina et al. (2014) (+)<br>Thornton and Vasilakis (2017) (+)<br>Vega and Winkelried (2005) (-) | Developing countries<br>Mixed sample<br>Mixed sample<br>Mixed sample<br>Mixed sample<br>Advanced countries<br>Developing countries<br>Mixed sample<br>Mixed sample |
| Exchange rate volatility | Samarina and De Haan (2014) (+)<br>Samarina and Sturm (2014) (+)<br>Hu (2006) (x) | Mixed sample<br>Mixed sample<br>Mixed sample |
| Trade openness | Arsić et al (2022) (-)<br>de Mendonça and de Guimarães e Souza (2012) (x)<br>Fouejieu (2017) (x)<br>Fry-McKibbin and Wang (2014) (x)<br>Fry-McKibbin and Wang (2014) (-)<br>Hu (2006) (x)<br>Leyva (2008) (+)<br>Lucotte (2010) (+)<br>Lucotte (2012) (x)<br>Lin (2010) (-)<br>Lin (2010) (x)<br>Lin and Ye (2007) (x)<br>Lin and Ye (2009) (-)<br>Minea et al. (2021) (-)<br>Minea and Tapsoba (2014) (-)<br>Mishkin and Schmidt-Hebbel (2002) (+)<br>Rose (2014) (x)<br>Samarina and De Haan (2014) (x)<br>Samarina et al. (2014) (+)<br>Samarina et al. (2014) (-)<br>Thornton and Vasilakis (2017) (+)<br>Thornton and Vasilakis (2017) (-)<br>Thornton and Vasilakis (2017) (-)<br>Vega and Winkelried (2005) (-)<br>Wang (2016) (x)<br>Yamada (2013) (x) | EMEs<br>Mixed sample<br><br>EMEs<br>Advanced countries<br>EMEs<br>Mixed sample<br>Mixed sample<br>EMEs<br>EMEs<br>Developing countries<br>Mixed sample<br>Advanced countries<br>Developing countries<br>Developing countries<br>Mixed sample<br>Mixed sample<br>Mixed sample<br>Mixed sample<br>Advanced countries<br>Developing countries<br>Mixed sample<br>Advanced countries<br>Developing countries<br>Mixed sample<br>EMEs<br>Developing countries |
| Financial openness | de Mendonça and de Guimarães e Souza (2012) (+)<br>Rose (2014) (x)<br>Samarina and De Haan (2014) (-)<br>Samarina et al. (2014) (x)<br>Samarina et al. (2014) (+)<br>Thornton and Vasilakis (2017) (+) | Mixed sample<br><br>Mixed sample<br>Mixed sample<br>Advanced countries<br>Developing countries<br>Mixed sample |
| | Thornton and Vasilakis (2017) (+) | Developing countries |
| Current account balance | Arsić et al (2022) (-)<br>Hu (2006) (x)<br>Mukherjee and Singer (2008) (+) | EMEs<br>Mixed sample<br>Mixed sample |
| Central bank independence | Fouejieu (2017) (+)<br>Hu (2006) (x)<br>Lin and Ye (2007) (+)<br>Lin and Ye (2009) (x)<br>Lucotte (2010) (+)<br>Lucotte (2012) (+)<br>Minea and Tapsoba (2014) (+) | EMEs<br>Mixed sample<br>Advanced countries<br>Developing countries<br>EMEs<br>EMEs<br>Mixed sample |
| Central bank instrument independence | Mishkin and Schmidt-Hebbel (2002) (+)<br>Samarina and De Haan (2014) (x)<br>Samarina and De Haan (2014) (x)<br>Samarina and De Haan (2014) (+) | Mixed sample<br>Mixed sample<br>Advanced countries<br>Developing countries |
| Central bank goal independence | Mishkin and Schmidt-Hebbel (2002) (-) | Mixed sample |
| Legal central bank independence | Mishkin and Schmidt-Hebbel (2002) (x) | Mixed sample |
| Central bank without bank regulatory authority | Mukherjee and Singer (2008) (+) | Mixed sample |
| Central bank restrictions on government lending | Carare and Stone (2006) (x)<br>Carare and Stone (2006) (-) | Mixed sample<br>EMEs |
| Monetary targets | Mishkin and Schmidt-Hebbel (2002) (-) | Mixed sample |
| Fiscal balance | Carare and Stone (2006) (x)<br>de Mendonça and de Guimarães e Souza (2012) (-)<br>Hu (2006) (+)<br>Leyva (2008) (x)<br>Lin (2010) (+)<br>Lin and Ye (2007) (x)<br>Mishkin and Schmidt-Hebbel (2002) (x)<br>Pontines (2013) (-)<br>Samarina and De Haan (2014) (x)<br>Vega and Winkelried (2005) (x) | Mixed sample<br>Mixed sample<br>Mixed sample<br>Mixed sample<br>Mixed sample<br>Advanced countries<br>Mixed sample<br>Mixed sample<br>Mixed sample<br>Mixed sample |
| Government debt | Arsić et al (2022) (+)<br>Carare and Stone (2006) (x)<br>Carare and Stone (2006) (x)<br>Ismailov et al. (2016) (-)<br>Ismailov et al. (2016) (x)<br>Ismailov et al. (2016) (-)<br>Lin and Ye (2009) (x)<br>Minea and Tapsoba (2014) (-)<br>Samarina and De Haan (2014) (-)<br>Samarina and De Haan (2014) (-) | EMEs<br>Mixed sample<br>EMEs<br>Mixed sample<br>High-income countries<br>Low-income countries<br>Developing countries<br>Mixed sample<br>Mixed sample<br>Advanced countries |
| | Samarina and De Haan (2014) (x)<br>Samarina et al. (2014) (x)<br>Samarina et al. (2014) (-)<br>Samarina and Sturm (2014) (x)<br>Thornton and Vasilakis (2017) (-)<br>Wang (2016) (x) | Developing countries<br>Advanced countries<br>Developing countries<br>Mixed sample<br>Mixed sample<br>EMEs |
| External debt | Samarina and De Haan (2014) (x)<br>Hu (2006) (-) | Mixed sample<br>Mixed sample |
| Tax revenue | Minea et al. (2021) (-) | Developing countries |
| Central bank's assets | Ardakani et al. (2018) (+)<br>Ardakani et al. (2018) (+)<br>Ardakani et al. (2018) (-) | Mixed sample<br>Advanced countries<br>Developing countries |
| Foreign reserve | Yamada (2013) (x) | Developing countries |
| Financial development | Ardakani et al. (2018) (-)<br>Ardakani et al. (2018) (-)<br>Ardakani et al. (2018) (-)<br>Carare and Stone (2006) (+)<br>de Mendonça and de Guimarães e Souza (2012) (+)<br>Hu (2006) (x)<br>Leyva (2008) (+)<br>Lucotte (2010) (x)<br>Lucotte (2012) (x)<br>Samarina and De Haan (2014) (-)<br>Samarina and De Haan (2014) (-)<br>Samarina and De Haan (2014) (x)<br>Samarina and Sturm (2014) (+)<br>Samarina et al. (2014) (+)<br>Thornton and Vasilakis (2017) (+)<br>Vega and Winkelried (2005) (+) | Mixed sample<br>Advanced countries<br>Developing countries<br>Mixed sample<br>Mixed sample<br>Mixed sample<br>Mixed sample<br>EMEs<br>EMEs<br>Mixed sample<br>Advanced countries<br>Developing countries<br>Mixed sample<br>Advanced countries<br>Mixed sample<br>Mixed sample |
| Financial structure | Samarina and De Haan (2014) (x) | Mixed sample |
| Financial crises/ Financial instability | Samarina and De Haan (2014) (x)<br>Thornton and Vasilakis (2017) (x)<br>Thornton and Vasilakis (2017) (-) | Mixed sample<br>Advanced countries<br>Developing countries |
| Institutional quality | Minea et al. (2021) (-) | Developing countries |
| Constraints on the executive | Minea et al. (2021) (+) | Developing countries |
| Right-wing government | Mukherjee and Singer (2008) (+) | Mixed sample |
| Democracy | Rose (2014) (+) | Mixed sample |
| Political risk | Ismailov et al. (2016) (x) | Mixed sample |
| Country size / population size | Arsić et al (2022) (+)<br>Rose (2014) (+)<br>Wang (2016) (+)<br>Yamada (2013) (+) | EMEs<br>Mixed sample<br>EMEs<br>Developing countries |
| Government and overall political stability | Lucotte (2010) (+) | EMEs |
| Political polarization | Lucotte (2010) (+) | EMEs |
| The number of veto players | Lucotte (2010) (+) | EMEs |
| Federalism / Decentralization | Lucotte (2010) (+) | EMEs |
| Number of inflation targeters / neighboring inflation targeters | Lucotte (2012) (+)<br>Wang (2016) (x)<br>Yamada (2013) (+) | EMEs<br>EMEs<br>Developing countries |
| Primary schooling | Minea et al. (2021) (+) | Developing countries |

Note: "+" and "-" indicate higher and lower likelihood to adopt IT, respectively, while "x" indicates that the factor is either statistically insignificant or economically negligible.

Table 3A. Selected empirical studies on the macroeconomic effects of IT

| Study | Data and methodology | Findings |
|---|---|---|
| Abo-Zaid and Tuzemen (2012) | 50 advanced and developing countries during 1980-2007; difference-in-difference | inflation (x) in advanced countries, and (-) in developing countries,<br>inflation volatility (x) in both samples,<br>GDP growth (+) in both samples,<br>GDP growth volatility (x) in advanced countries and (-) in developing countries,<br>fiscal balance (+) in advanced countries and (x) in developing countries. |
| Almeida and Goodhart (1998) | 7 OECD countries during 1986-1997; (E)GARCH | interest rate volatility (-),<br>exchange rate volatility (x). |
| Alpanda and Honig (2014) | 66 advanced countries and EMEs during 1980-2006; GMM and difference-in-difference | inflation (x) in both samples. |
| Amira et al. (2013) | 36 EMEs during 1979-2009; OLS and GMM | GPP growth (+),<br>GDP growth volatility (+). |
| Angeriz and Arestis (2007a) | 7 OECD countries and EMEs during 1994-2005; intervention analysis to multivariate time series models | inflation (x),<br>inflation volatility (x),<br>inflation expectations (x). |
| Arestis et al. (2014) | 22 OECD during 1990-2011; time series analysis | inflation (x). |
| Ardakani et al. (2018) | 98 advanced and developing countries during 1998-2013; propensity score matching | inflation (x) in both samples,<br>inflation volatility (x) in both samples,<br>interest rate volatility (x) in both samples,<br>public debt (-) in both samples,<br>real exchange rate volatility (+) in advanced countries,<br>real exchange rate (x) in developing countries,<br>sacrifice ratio (-) in advanced countries. |
| Arsić et al. (2022) | 26 EMEs during 1997–2019; dynamic panel model and propensity score matching | inflation (-),<br>inflation volatility (-),<br>inflation persistence (x), |
| | | GDP growth (x),<br>GDP volatility (-). |
| Ayres et al. (2014) | 51 developing countries during 1985-2010; OLS and fixed-effects | inflation (x),<br>GDP growth (x). |
| Ball (2010) | 20 OECD countries over 1985-2007; difference-in-difference | inflation (-),<br>GDP growth (x),<br>GDP growth volatility (x),<br>long-term interest rate (+),<br>long-term interest rate volatility (+). |
| Ball and Sheridan (2004) | 20 OECD countries during 1960-2001; difference-in-difference | inflation (x),<br>inflation volatility (x),<br>inflation persistence (x),<br>inflation expectations (x),<br>GDP growth (x),<br>output volatility (x),<br>long-term interest rate (x), short-term interest rate volatility (x). |
| Barnebeck Andersen et al. (2015) | 196 advanced and developing countries during 2007-2013; OLS | GDP growth (+) in both samples. |
| Batini and Laxton (2007) | 42 EMEs during 1990-2004; difference-in-difference | inflation (-),<br>inflation volatility (-),<br>inflation expectations (-),<br>output volatility (x),<br>real interest rate volatility (-),<br>nominal exchange rate (-). |
| Baxa et al. (2014) | 5 OECD countries during 1975(1981)-2007; moment-based estimator | inflation persistence (-). |
| Baxa et al. (2015) | 3 OECD countries during 1996-2010; Bayesian model averaging | inflation persistence (x),<br>inflation volatility (-). |
| Benati (2008) | 5 advanced countries during 1834-2005; grid bootstrap median-unbiased estimator and Bayesian DSGE model | inflation persistence (-). |
| Berganza and Broto (2012) | 37 EMEs during 1995-2010; OLS | real exchange rate volatility (+). |
| Berument and Yuksel (2007) | 5 OECD countries and 4 EMEs; GARCH | inflation (x),<br>inflation volatility (x). |
| Bratsiotis et al. (2015) | 7 advanced countries during 1946-2001; OLS and time series analysis | inflation persistence (-). |
| Brito and Bystedt (2010) | 46 developing countries during 1980-2006; GMM | inflation (-),<br>GDP growth (-),<br>inflation volatility (x),<br>output volatility (x). |
| Broz and Plouffe (2010) | 10 090 firms from 81 countries in 2001; ordered probit | inflation expectations (x). |
| Bundick and Smith (2018) | Japan during 2001-2017; OLS | inflation expectations (x). |
| Calderón and Schmidt-Hebbel (2010) | 97 advanced and developing countries during 1975-2005; IV estimation with fixed- and random-effects models, pooled mean group estimator, and GMM. | inflation (-). |
| Canarella and Miller (2016) | 11 advanced countries and EMEs during 1976-2013; fractional integration and time series analysis | inflation persistence (x). |
| Canarella and Miller (2017a) | 13 OECD countries during 1976-2013; fractional integration | inflation persistence (x). |
| Canarella and Miller (2017b) | 6 OECD countries during 1976-2013; fractional integration | inflation persistence (x) . |
| Capistrán and Ramos-Francia (2010) | 25 advanced and developing countries during 1989-2006; fixed-effects | inflation expectations (x) in advanced countries and (-) in developing countries. |
| Castelnuovo et al. (2003) | 15 advanced countries during 1990-2002; time series analysis | inflation expectations (-). |
| Cecchetti and Hakkio (2010) | 15 advanced countries during 1989-2009; OLS, SUR, fixed- and random-effects | inflation expectations (x). |
| Chadha and Nolan (2001) | UK during 1987-1999; state-space model | interest-rate volatility (+). |
| Chevapatrakul and Paez-Farrell (2018) | 40 EMEs during 1980-2014; quantile regression | inflation (x),<br>inflation volatility (x),<br>output volatility (x). |
| Chiquiar et al. (2010) | Mexico during 1995-2006; time series analysis | inflation persistence (-). |
| Choi et al. (2003) | New Zealand during 1982-1996; Markov switching model | inflation volatility (-),<br>GDP growth volatility (-). |
| Çiçek and Akar (2013) | Turkey during 1994-2012; quantile autoregression | inflation persistence (-). |
| Combes et al. (2014) | 152 advanced and developing countries during 1990-2009; GMM | inflation (-),<br>budget balance (+). |
| Corbo et al. (2001) | 26 advanced countries and EMEs during 1980-1999; VAR | inflation expectations (-). |
| Corbo et al. (2002) | 23 advanced countries and EMEs during 1980-1999; VAR | inflation expectations (-),<br>inflation persistence (-). |
| Crowe (2010) | 11 advanced countries and EMEs; OLS, 2SLS, propensity score matching and difference-in-difference | inflation expectations (x). |
| Davis and Presno (2014) | 36 advanced and developing countries during 1990-2011; SVAR | inflation expectations (x) in advanced and (-) in developing countries,<br>inflation persistence (x) in advanced and (-) in developing countries. |
| de Carvalho Filho (2010) | 84 advanced and developing countries during 2002-2009; fixed-effects | nominal interest rate (-),<br>real interest rate (-),<br>real exchange rate (-)<br>unemployment rate (x),<br>GDP growth (+) in advanced countries and (x) in developing countries. |
| de Carvalho Filho (2011) | 51 advanced countries and EMEs during 2006-2010; OLS, fixed-effects, and Bayesian model averaging | GDP growth (+) in both samples. |
| de Mendonça (2007) | 14 OECD countries; OLS | inflation (x),<br>GDP growth (+),<br>interest rate (-). |
| de Mendonça and de Guimarães e Souza (2012) | 180 advanced and developing countries during 1990-2007; propensity score matching | inflation (x) in advanced countries and (-) in EMEs,<br>inflation volatility (x) in advanced countries and (-) in EMEs. |
| De Pooter et al. (2014) | 3 EMVEs during 2002-2013; OLS | inflation expectations (-). |
| Debelle (1996) | 3 advanced countries during 1975-1995; time series analysis | inflation expectations (x). |
| Demertzis et al. (2010) | 7 advanced countries during 1999-2008; VAR | inflation expectations (-). |
| Dotsey (2006) | 11 advanced countries during 1979-2004; descriptive statistics | inflation (-),<br>inflation volatility (-),<br>GDP growth (+),<br>GDP growth volatility (-). |
| Dueker and Fischer (2006) | 6 advanced economies during 1970-2005; two-state Markov switching model | inflation (x). |
| Duong (2021) | 54 EMEs during 2002-2010; difference-in-difference | inflation (x),<br>GDP growth (x). |
| Edwards (2007) | 7 advanced countries and EMEs during 1985-2005; SUR and GARCH | inflation persistence (x),<br>exchange rate volatility (x). |
| Ehrmann (2015) | 15 advanced countries during 1990-2014; OLS | inflation expectations (-). |
| Filardo and Genberg (2009) | 12 advanced countries and EMEs during 1985-2008; time-series analysis, OLS, and panel-data model | inflation volatility (x),<br>inflation persistence (x),<br>inflation expectations (x). |
| Fouejieu (2013) | 79 advanced and developing countries during 2003-2009; difference-in-difference | inflation (x),<br>inflation volatility (-),<br>GDP growth (x),<br>nominal and real interest rates (-). |
| Fratzscher et al. (2020) | 76 advanced and developing countries during 1970-2015; dynamic panel data model | inflation (-) in advanced countries,<br>GDP growth (+) in advanced countries,<br>inflation volatility (-) in both samples,<br>short-term interest rate (+) in advanced countries. |
| Freeman and Willis (1995) | 4 advanced countries during 1984-1993; time series analysis and OLS | inflation expectations (x),<br>real long-term interest rate (+),<br>nominal long-term interest rate volatility (x). |
| Fry-McKibbin and Wang (2014) | 31 developed and 60 EMEs during 2007-2012; propensity score matching | GDP growth (+) in advanced countries and (x) in EMEs,<br>inflation (+) in advanced countries and (x) in EMEs,<br>tax revenue (-) in both samples,<br>public debt (-) in both samples. |
| Ftiti and Hichri (2014) | 6 advanced countries during 1972-2008; time series analysis | inflation (-). |
| Gadea and Mayoral (2006) | 21 OECD countries during 1957-2003; fractional integration | inflation persistence (x). |
| Gagnon (2013) | 172 advanced and developing countries during 2007-2012; OLS | GDP growth (x),<br>inflation volatility (-). |
| Gemayel et al. (2011) | 57 EMEs during 1990-2008; difference-in-difference, OLS, fixed-effects, GMM | inflation (-),<br>inflation volatility (-),<br>GDP growth (x),<br>GDP growth volatility (x). |
| Genc and Balcilar (2012) | Turkey during 1994-2006; ARMA and regime-switching model | inflation (x). |
| Genc et al. (2007) | 4 advanced countries during 1960-2004; ARMA, GARCH, and regime switching model | inflation (x). |
| Gerlach and Tillmann (2012) | 19 advanced and EMEs during 1985-2010; AR model | inflation persistence (x). |
| Gillitzer and Simon (2015) | Australia, ove 1991-2013, OLS regression | inflation expectations (-). |
| Ginindza and Maasoumi (2013) | 30 advanced countries during 1980-2007; hyperbolic mean score function | inflation (-),<br>inflation volatility (x). |
| Gonçalves and Salles (2008) | 36 EMEs during 1980-2005; difference-in-difference | inflation (-),<br>inflation volatility (x),<br>output volatility (-). |
| Gupta et al. (2017) | South Africa during 1975-2015; quantile regression | inflation persistence (-). |
| Gürkaynak et al. (2007) | 2 advanced countries and 1 EME during 1994-2005; OLS | inflation expectations (-). |
| Gürkaynak et al. (2010) | 3 advanced countries during 1993-2005; OLS | inflation expectations (-). |
| Hossain (2014) | Australia during 1950-2010; time series analysis | inflation persistence (x),<br>inflation volatility (x). |
| Hu (2006) | 66 advanced and developing countries during 1980-2000; OLS | inflation (-),<br>inflation variability (x),<br>GDP growth (+),<br>GDP growth volatility (-). |
| Huh (1996) | UK during 1973-1995; VAR | inflation expectations (-),<br>short-term interest rate (-),<br>long-term interest rate (-). |
| IMF (2006) | 42 EMEs during 1990s-2000s; descriptive statistics | inflation (-),<br>inflation variability (-), |
| | | inflation expectations (-),<br>output volatility (-),<br>real interest rate volatility (-),<br>nominal exchange rate (-). |
| Johnson (2002) | 11 developed countries during 1984 to 2000; panel data model | inflation expectations (-). |
| Johnson (2003) | 5 advanced countries during 1984-1998; OLS | inflation expectations (-). |
| Kaseeram and Contogiannis (2011) | South Africa during 1960-2010; GARCH and GARCH-M, | inflation persistence (x),<br>inflation volatility (x). |
| Kočenda and Varga (2018) | 68 advanced countries and EMEs during 1993-2013; time series analysis and panel data model | inflation persistence (-). |
| Kontonikas (2004) | GARCH, GARCH-M, T-GARCH models, UK, 1972-2002. | inflation volatility (-),<br>inflation persistence (-). |
| Kose et al. (2018) | 37 advanced countries and EMEs during 1996-2015; difference-in-difference | inflation (-),<br>inflation volatility (-),<br>GDP growth (x),<br>real exchange rate (x),<br>real exchange rate volatility (x). |
| Kuttner and Posen (1999) | 3 advanced countries during 1982-1998; time series analysis | inflation persistence (x). |
| Kuttner and Posen (2001) | 41 advanced countries and EMEs during 1973-2000; time-series analysis, OLS and weighted least squares | nominal exchange rate volatility (x),<br>inflation persistence (-),<br>inflation (-). |
| Lane and Van Den Heuvel (1998) | UK during 1975-1997; Bayesian VAR | inflation (x),<br>inflation expectations (-),<br>short-term interest rate (-),<br>long-term interest rate (-),<br>GDP growth (x),<br>exchange rate (x). |
| Laubach and Posen (1997) | 8 advanced countries during 1971-1995; VAR | inflation (x),<br>GDP growth (x),<br>real interest rate (-). |
| Lee (1999) | 6 advanced countries during 1975-1996; cointegration, VAR, and OLS | GDP growth (x),<br>inflation (x),<br>short- and long-term interest rates (x). |
| Lee (2010) | 11 advanced countries during 1995-2007; OLS | short-term interest rate (+). |
| Lee (2011) | 60 EMEs during1993-2006; synthetic control methods | inflation (x). |
| Levin et al. (2004) | 12 advanced countries during 1994-2003 and five EMEs during 1990s; OLS and event-study analysis | inflation expectations (-) in advanced countries and (x) in EMEs,<br>inflation persistence (-) in advanced countries. |
| Levin and Piger (2004) | 12 advanced countries during 1984-2003; median unbiased estimator and time-series analysis | inflation persistence (x). |
| Lin (2010) | 74 advanced countries and EMEs during 1985-2005; propensity score matching | nominal and real exchange rate volatility (+) in advanced countries and (-) in developing countries,<br>foreign reserves (-) in advanced countries and (+) in developing countries,<br>current account (x) in both samples. |
| Lin and Ye (2007) | 22 advanced countries during 1985-1999; propensity score matching | inflation (x),<br>inflation volatility (x),<br>long-term nominal interest rate (x),<br>long-term nominal interest rate volatility (x). |
| Lin and Ye (2009) | 52 developing countries during 1985-2005; propensity score matching | inflation (-) EMEs,<br>inflation volatility (-) EMEs., |
| Lucotte (2012) | 30 EMEs during 1980-2004; propensity score matching | public revenue (+). |
| Miles (2007) | 17 EMEs since 1989; OLS | government consumption (+). |
| Minea and Tapsoba (2014) | 84 advanced and developing countries during 1985-2007; propensity score matching | budget balance (+) in the whole sample, (x) in advanced countries, and (+) in developing countries. |
| Minea et al. (2021) | 53 developing countries during 1984-2007; propensity score matching | public revenue (+). |
| Mollick et al. (2011) | 55 advanced countries and EMEs during 1986-2004; random-effects and GMM | GDP growth (+) in both samples. |
| Mukherjee and Singer (2008) | 49 OECD and non-OECD countries during 1987-2003; propensity score matching | inflation (-) in both samples. |
| Nadal-De Simone (2001) | 12 advanced countries during 1976-2000; time-varying parameter state-space model | output volatility (x). |
| Naqvi and Rizvo (2009) | 10 advanced countries and EMEs during 1987-2007; difference-in-difference | inflation (x),<br>inflation volatility (x),<br>GDP growth (x),<br>GDP growth volatility (x),<br>output gap volatility (x),<br>short-term interest rate volatility (x). |
| Neumann and von Hagen (2002) | 9 advanced countries during 1978-2001; difference-in-difference | inflation (x),<br>inflation volatility (x),<br>long- and short-term interest rates (-). |
| Ouyang and Rajan (2016) | 31 developing countries during 2007-2012; OLS | real exchange rate volatility (x),<br>inflation (-)<br>GDP growth (x). |
| Ouyang et al. (2016) | 62 advanced and developing countries during 2006-2012; propensity score matching, random-effects, and GMM | real exchange rate volatility (+) in advanced countries and (x) in developing countries. |
| Parkin (2014) | 26 advanced countries during 1980-2011; t-test of the difference across samples | inflation (-),<br>inflation volatility (-),<br>GDP growth volatility (-),<br>GDP growth (+). |
| Pétursson (2004) | 29 advanced countries and EMEs during 1981-2002; SUR, panel-data model, and VAR | inflation (x),<br>inflation persistence (-),<br>GDP growth (x),<br>short-run nominal interest rate (-). |
| Phiri (2016) | 46 developing countries during 1994-2014; OLS, fixed-effects, and random-effects | inflation persistence (-). |
| Pontines (2013) | 74 advanced and developing countries during 1985-2005; treatment effect regression | nominal and real exchange rate volatility (+) in advanced countries and (-) in developing countries. |
| Porter and Yao (2005) | Mauritius during 1996-2004; maximum likelihood | inflation (-),<br>inflation expectations (-). |
| Ravenna (2007) | Canada during 1991-2007; Kalman filter and DSGE model | inflation (x). |
| Roger (2009) | 84 advanced and developing countries during 1990s-2008; descriptive analysis | inflation (-) in both low-income and high-income countries,<br>inflation volatility (-) in low-income countries,<br>GDP growth volatility (-) in low-income countries,<br>GDP growth (x) in low-income countries and (+) in high-income countries. |
| Roger (2010) | 26 advanced and developing countries during 1991-2009; descriptive analysis | inflation (-) in low-income countries,<br>GDP growth (+) in low-income countries and (x) in high-income countries,<br>inflation volatility (-) in low-income countries,<br>GDP growth volatility (-) in low-income countries and (x) in high-income countries. |
| Rose (2007) | 45 advanced and developing countries during 1990-2005; OLS | nominal and real exchange rate volatility (x),<br>foreign reserves (x),<br>current account (x). |
| Rose (2014) | 170 advanced and developing countries during 2007-2012; panel data model and IV | capital flows (x)<br>current account (x),<br>foreign reserves (x),<br>broad money (x),<br>fiscal policy (x),<br>inflation (x),<br>real exchange rate (x),<br>property prices (x),<br>bond yields (x). |
| Ryczkowski and Ręklewski (2021) | 27 OECD countries during 2000-2017; OLS, fixed-effects, random-effects, between-effects, and Hausman-Taylor model | GDP growth (+). |
| Samarina et al. (2014) | 58 advanced and developing countries during 1985-2011; difference-in-difference and propensity score matching | inflation (x) in advanced countries and (-) in developing countries. |
| Schmidt-Hebbel and Werner (2002) | 3 EMEs during 1986-2001; VAR. | inflation (x),<br>inflation expectations (-). |
| Siklos (1999) | 10 advanced countries during 1958-1997; time series analysis | inflation persistence (x). |
| Siklos (2008) | 29 advanced countries and EMEs during 1993-2005; time series analysis | inflation persistence (x). |
| Siregar and Goo (2010) | 2 EMEs during 1990-2008; ARDL model | inflation persistence (x). |
| Stojanovikj and Petrevski (2021) | 44 EMEs during 1970-2017; GMM | inflation (-),<br>inflation variability (+). |
| Suh and Kim (2021) | 29 advanced countries and EMEs during 1979-2018; panel-data model | inflation expectations (-) in both samples. |
| Thornton (2016) | 72 developing countries during 1980-2005; difference-in-difference | inflation (x),<br>GDP growth volatility (x). |
| Vega and Winkelried (2005) | 91 advanced and developing countries during 1990-2004; propensity score matching | inflation (-) in both samples,<br>inflation volatility (-) in both samples,<br>inflation persistence (-) in both samples. |
| Walsh (2009) | 22 advanced countries during 1985-1999; propensity score matching | GDP growth (x),<br>GDP growth volatility (x). |
| Wang (2016) | 16 EMEs during 1990-2010; propensity score matching | inflation (x),<br>inflation volatility (x). |
| Willard (2012) | 22 OECD countries during 1985-2002; OLS, IV, GMM, structural model, and difference-in-difference | inflation (x),<br>inflation volatility (x),<br>inflation expectations (x),<br>inflation persistence (x). |
| Wu (2004) | 22 OECD countries during 1985-2002; difference-in-difference | inflation (-),<br>real interest rate (x). |
| Yamada (2013) | 121 developing countries during 1995-2007; propensity score matching | inflation (-). |
| Yigit (2010) | 8 advanced countries during 1961-2009; fractional integration | inflation expectations (x),<br>inflation persistence (x). |

Note: "+" indicates that IT is associated with higher value of the specified macroeconomic variable (e.g., inflation, output, debt etc.); "-" indicates that IT is associated with lower value of the macroeconomic variable; "x" indicates statistically insignificant, economically negligible or mixed (non-robust) evidence.

Table 3B. Summary of the empirical evidence on macroeconomic effects of inflation targeting

| Variable | Findings | Sample |
|---|---|---|
| Inflation | Abo-Zaid and Tuzemen (2012) (x) | Advanced countries |
| | Abo-Zaid and Tuzemen (2012) (-) | Developing countries |
| | Almeida and Goodhart (1998) (x) | Advanced countries |
| | Alpanda and Honig (2014) (x) | Mixed sample |
| | Angeriz and Arestis (2007) (x) | Advanced countries |
| | Angeriz and Arestis (2007) (x) | EMEs |
| | Ardakani et al. (2018) (x) | Mixed sample |
| | Arestis et al. (2014) (x) | Advanced countries |
| | Arsić et al. (2022) (-) | EMEs |
| | Ayres et al. (2014) (x) | Mixed sample |
| | Ball (2010) (-) | Advanced countries |
| | Ball and Sheridan (2004) (x) | Advanced countries |
| | Batini and Laxton (2007) (-) | EMEs |
| | Berument and Yuksel (2007) (x) | Advanced countries |
| | Brito and Bystedt (2010) (-) | EMEs |
| | Calderón and Schmidt-Hebbel (2010) (-) | Mixed sample |
| | Chevapatrakul and Paez-Farrell (2018) (x) | EMEs |
| | Combes et al. (2014) (-) | Mixed sample |
| | De Mendonça (2007) (x) | Advanced countries |
| | de Mendonça and de Guimarães e Souza (2012) (-) | EMEs |
| | de Mendonça and de Guimarães e Souza (2012) (x) | Advanced countries |
| | Dotsey (2006) (-) | Advanced countries |
| | Gonçalves and Salles (2008) (-) | EMEs |
| | Dueker and Fischer (1996) (x) | Advanced countries |
| | Duong (2021) (x) | EMEs |
| | Fouejieu (2013) (x) | Advanced countries |
| | Fouejieu (2013) (x) | EMEs |
| | Fratzscher et al. (2020) (-) | Advanced countries |
| | Fry-McKibbin and Wang (2014) (+) | Advanced countries |
| | Fry-McKibbin and Wang (2014) (x) | EMEs |
| | Ftiti and Hichri (2014) (-) | Advanced countries |
| | Gemayel et al. (2011) (-) | EMEs |
| | Genc and Balcilar (2012) (x) | Turkey |
| | Genc et al. (2007) (x) | Advanced countries |
| | Ginindza and Maasoumi (2013) (-) | Advanced countries |
| | Hu (2006) (-) | Mixed sample |
| | IMF (2006) (-) | EMEs |
| | Kose et al. (2018) (-) | Mixed sample |
| | Kuttner and Posen (2001) (-) | Mixed sample |
| | Lane and Van Den Heuvel (1998) (x) | UK |
| | Laubach and Posen (1997) (x) | Advanced countries |
| | Lee (1999) (x) | Advanced countries |
| | Lee (2011) (x) | EMEs |
| | Lin and Ye (2007) (x) | Advanced countries |
| | Lin and Ye (2009) (-)<br>Mukherjee and Singer (2008) (-)<br>Naqvi and Rizvo (2009) (x)<br>Neumann and von Hagen (2002) (x)<br>Ouyang and Rajan (2016) (-)<br>Parkin (2014) (-)<br>Pétursson (2004) (x)<br>Porter and Yao (2005) -)<br>Ravenna (2007) (x)<br>Roger (2009) (-)<br>Roger (2009) (-)<br>Roger (2010) (-)<br>Rose (2014) (x)<br>Samarina et al. (2014) (x)<br>Samarina et al. (2014) (-)<br>Schmidt-Hebbel and Werner (2002) (x)<br>Stojanovikj and Petrevski (2021) (-)<br>Thornton (2016) (x)<br>Vega and Winkelried (2005) (-)<br>Wang (2016) (x)<br>Willard (2012) (x)<br>Wu (2004) (-)<br>Yamada (2013) (-) | EMEs<br>Mixed sample<br>Mixed sample<br>Advanced countries<br>EMEs<br>Advanced countries<br>Mixed sample<br>Mauritius<br>Canada<br>Low-income countries<br>High-income countries<br>Low-income countries<br>Mixed sample<br>Advanced countries<br>EMEs<br>Brazil, Chile, and Mexico<br>EMEs<br>EMEs<br>Mixed sample<br>EMEs<br>Advanced countries<br>Advanced countries<br>Developing countries |
| Inflation volatility | Abo-Zaid and Tuzemen (2012) (x)<br>Almeida and Goodhart (1998) (+)<br>Angeriz and Arestis (2007) (x)<br>Ardakani et al. (2018) (x)<br>Arsić et al. (2022) (-)<br>Ball and Sheridan (2004) (x)<br>Batini and Laxton (2007) (-)<br>Baxa et al. (2015) (-)<br>Berument and Yuksel (2007) (x)<br>Brito and Bystedt (2010) (x)<br>Chevapatrakul and Paez-Farrell (2018) (x)<br>Choi et al. (2003) (-)<br>de Mendonça and de Guimarães e Souza (2012) (-)<br>de Mendonça and de Guimarães e Souza (2012) (x)<br>Dotsey (2006) (-)<br>Filardo and Genberg (2009) (x)<br>Fouejieu (2013) (-)<br>Fratzscher et al. (2020) (-)<br>Gagnon (2013) (-)<br>Gemayel et al. (2011) (-)<br>Ginindza and Maasoumi (2013) (x)<br>Gonçalves and Salles (2008) (x)<br>Hossain (2014) (x) | Mixed sample<br>Advanced countries<br>Mixed sample<br>Mixed sample<br>EMEs<br>Advanced countries<br>EMEs<br>Advanced countries<br>Advanced countries<br>EMEs<br>EMEs<br>New Zealand<br>EMEs<br>Advanced countries<br>Advanced countries<br>Mixed sample<br>Mixed sample<br>Mixed sample<br>Mixed sample<br>EMEs<br>Advanced countries<br>EMes<br>Australia |
| | Hu (2006) (x)<br>IMF (2006) (-)<br>Kaseeram and Contogiannis (2011) (x)<br>Kontonikas (2004) (-)<br>Kose et al. (2018) (-)<br>Lee (2010) (-)<br>Lin and Ye (2007) (x)<br>Lin and Ye (2009) (-)<br>Naqvi and Rizvo (2009) (x)<br>Neumann and von Hagen (2002) (x)<br>Parkin (2014) advanced (-)<br>Roger (2009) (-)<br>Roger (2010) (-)<br>Stojanovikj and Petrevski (2021) (+)<br>Vega and Winkelried (2005) (-)<br>Wang (2016) (x)<br>Willard (2012) (x) | Mixed sample<br>EMEs<br>South Africa<br>UK<br>Mixed sample<br>EMEs<br>Advanced countries<br>EMEs<br>Mixed sample<br>Advanced countries<br>Advanced countries<br>Low-income countries<br>Low-income countries<br>EMEs<br>Mixed sample<br>EMEs<br>Advanced countries |
| Inflation expectations and inflation persistence | Angeriz and Arestis (2007) (x)<br>Arsić et al. (2022) (x)<br>Ball and Sheridan (2004) (x)<br>Batini and Laxton (2007) (-)<br>Baxa et al. (2014) (-)<br>Baxa et al. (2015) (x)<br>Benati (2008) (-)<br>Bratsiotis et al. (2015) (-)<br>Broz and Plouffe (2010) (x)<br>Bundick and Smith (2018) (x)<br>Canarella and Miller (2016) (x)<br>Canarella and Miller (2017a) (x)<br>Canarella and Miller (2017a) (-)<br>Canarella and Miller (2017b) (x)<br>Capistrán and Ramos-Francia (2010) (-)<br>Capistrán and Ramos-Francia (2010) (x)<br>Castelnuovo et al. (2003) (-)<br>Cecchetti and Hakkio (2010) (x)<br>Chiquiar et al. (2010) (-)<br>Çiçek and Akar (2013) (-)<br>Corbo et al. (2001) (-)<br>IMF (2006) (-)<br>Corbo et al. (2002) (-)<br>Crowe (2010) (x)<br>Davis and Presno (2014) (x)<br>Davis and Presno (2014) (-)<br>De Pooter et al. (2014) (-)<br>Debelle (1996) (x)<br>Demertzis et al. (2010) (-) | Mixed sample<br>EMEs<br>Advanced countries<br>EMEs<br>Advanced countries<br>Advanced countries<br>Advanced countries<br>Advanced countries<br>Mixed sample<br>Japan<br>Advanced countries<br>Advanced countries<br>EMEs<br>Advanced countries<br>EMEs<br>Advanced countries<br>Advanced countries<br>Advanced countries<br>Mexico<br>Turkey<br>Mixed sample<br>EMEs<br>Mixed sample<br>Mixed sample<br>Advanced countries<br>Developing countries<br>EMEs<br>Advanced countries<br>Advanced countries |
| | Dueker and Fischer (2006) (x)<br>Edwards (2007) (x)<br>Ehrmann (2015) (-)<br>Filardo and Genberg (2009) (x)<br>Freeman and Willis (1995) (x)<br>Gadea and Mayoral (2006) (x)<br>Gerlach and Tillmann (2012) (x)<br>Gillitzer and Simon (2015) (-)<br>Gupta et al. (2017) (-)<br>Gürkaynak et al. (2007) (-)<br>Gürkaynak et al. (2010) (-)<br>Hossain (2014) (x)<br>Huh (1996) (-)<br>Johnson (2002) (-)<br>Johnson (2003) (-)<br>Kaseeram and Contogiannis (2011) (x)<br>Kočenda and Varga (2018) (-)<br>Kontonikas (2004) (-)<br>Kuttner and Posen (1999) (x)<br>Kuttner and Posen (2001) (-)<br>Lane and Van Den Heuvel (1998) (-)<br>Levin et al. (2004) (-)<br>Levin et al. (2004) (x)<br>Levin and Piger (2004) (x)<br>Neumann and von Hagen (2002) (x)<br>Pétursson (2004) (-)<br>Phiri (2016) (-)<br>Porter and Yao (2005) (-)<br>Schmidt-Hebbel and Werner (2002) (-)<br>Siklos (1999) (x)<br>Siklos (2008) (x)<br>Siregar and Goo (2010) (x)<br>Suh and Kim (2021) (-)<br>Vega and Winkelried (2005) (-)<br>Willard (2012) (x)<br>Yigit (2010) (x) | Advanced countries<br>Mixed sample<br>Advanced countries<br>Mixed sample<br>Advanced countries<br>Advanced countries<br>Mixed sample<br>Australia<br>South Africa<br>Canada, Chile, USA<br>Sweden, UK, USA<br>Australia<br>UK<br>Advanced countries<br>Advanced countries<br>South Africa<br>Mixed sample<br>UK<br>Canada, New Zealand, UK<br>Mixed sample<br>UK<br>Advanced countries<br>EMEs<br>Advanced countries<br>Advanced countries<br>Mixed sample<br>Developing countries<br>Mauritius<br>Brazil, Chile, Mexico<br>Advanced countries<br>Mixed sample<br>Indonesia, Thailand<br>Mixed sample<br>Mixed sample<br>Advanced countries<br>Advanced countries |
| GDP growth | Abo-Zaid and Tuzemen (2012) (+)<br>Amira et al. (2013) (+)<br>Ayres et al. (2014) (x)<br>Arsić et al. (2022) (x)<br>Ball (2010) (x)<br>Ball and Sheridan (2004) (x)<br>Barnebeck Andersen et al. (2015) (+)<br>Brito and Bystedt (2010) (-)<br>de Carvalho Filho (2010) (+)<br>de Carvalho Filho (2010) (x) | Mixed sample<br>EMEs<br>Developing countries<br>EMEs<br>Advanced countries<br>Advanced countries<br>Mixed sample<br>EMEs<br>Advanced countries<br>Developing countries |
| | de Carvalho Filho (2011) (+)<br>de Mendonça (2007) (+)<br>Dotsey (2006) (+)<br>Duong (2021) (x)<br>Fouejieu (2013) (x)<br>Fratzscher et al. (2020) (+)<br>Fry-McKibbin and Wang (2014) (+)<br>Fry-McKibbin and Wang (2014) (x)<br>Gagnon (2013) (x)<br>Gemayel et al. (2011) (x)<br>Hu (2006) (+)<br>Kose et al. (2018) (x)<br>Lane and Van Den Heuvel (1998) (x)<br>Laubach and Posen (1997) (x)<br>Lee (1999) (x)<br>Mollick et al. (2011) (+)<br>Naqvi and Rizvo (2009) (x)<br>Ouyang and Rajan (2016) (x)<br>Parkin (2014) (+)<br>Pétursson (2004) (x)<br>Roger (2009) (x)<br>Roger (2009) (+)<br>Roger (2010) (+)<br>Roger (2010) (x)<br>Ryczkowski and Ręklewski (2021) (+)<br>Walsh (2009) (x) | Mixed sample<br>Advanced countries<br>Advanced countries<br>EMEs<br>Mixed sample<br>Advanced countries<br>Advanced countries<br>EMEs<br>Mixed sample<br>EMEs<br>Mixed sample<br>Mixed sample<br>UK<br>Advanced countries<br>Advanced countries<br>Mixed sample<br>Mixed sample<br>EMEs<br>Advanced countries<br>Mixed sample<br>Low-income countries<br>High-income countries<br>Low-income countries<br>High-income countries<br>Advanced countries<br>Advanced countries |
| Output/growth volatility | Abo-Zaid and Tuzemen (2012) (-)<br>Abo-Zaid and Tuzemen (2012) (x)<br>Amira et al. (2013) (+)<br>Arsić et al. (2022) (-)<br>Ball (2010) (x)<br>Ball and Sheridan (2004) (x)<br>Batini and Laxton (2007) (x)<br>Brito and Bystedt (2010) (x)<br>Chevapatrakul and Paez-Farrell (2018) (x)<br>Choi et al. (2003) (-)<br>Dotsey (2006) (-)<br>Gemayel et al. (2011) (x)<br>Gonçalves and Salles (2008) (-)<br>Hu (2006) (-)<br>IMF (2006) (-)<br>Nadal-De Simone (2001) (x)<br>Naqvi and Rizvo (2009) (x)<br>Parkin (2014) (-)<br>Roger (2009) (-)<br>Roger (2010) (x) | Developing countries<br>Advanced countries<br>EMEs<br>EMEs<br>Advanced countries<br>Advanced countries<br>EMEs<br>EMEs<br>EMEs<br>New Zealand<br>Advanced countries<br>EMEs<br>EMEs<br>Mixed sample<br>EMEs<br>Advanced countries<br>Mixed sample<br>Advanced countries<br>Low-income countries<br>High-income countries |
| | Thornton (2016) (x)<br>Walsh (2009) (x) | EMEs<br>Advanced countries |
| Interest rate | Almeida and Goodhart (1998) (-)<br>Ball (2010) (+)<br>Ball and Sheridan (2004) (x)<br>Chadha and Nolan (2001) (+)<br>de Carvalho Filho (2010) (-)<br>de Mendonça (2007) (-)<br>Fouejieu (2013) (-)<br>Fratzscher et al. (2020) (+)<br>Freeman and Willis (1995) (+)<br>Huh (1996) (-)<br>Kose et al. (2018) (x)<br>Lane and Van Den Heuvel (1998) (-)<br>Laubach and Posen (1997) (-)<br>Lee (1999) (x)<br>Lee (2010) (+)<br>Lin and Ye (2007) (x)<br>Naqvi and Rizvo (2009) (x)<br>Neumann and von Hagen (2002) (-)<br>Pétursson (2004) (-)<br>Wu (2004) (x) | Advanced countries<br>Advanced countries<br>Advanced countries<br>UK<br>Mixed sample<br>Advanced countries<br>Mixed sample<br>Advanced countries<br>Advanced countries<br>UK<br>Mixed sample<br>UK<br>Advanced countries<br>Advanced countries<br>Advanced countries<br>Advanced countries<br>Mixed sample<br>Advanced countries<br>Mixed sample<br>Advanced countries |
| Interest rate volatility | Ardakani et al. (2018) (x)<br>Ardakani et al. (2018) (-)<br>IMF (2006) (-)<br>Batini and Laxton (2007) (-)<br>Freeman and Willis (1995) (x)<br>Kose et al. (2018) (x) | Advanced countries<br>Developing countries<br>EMEs<br>Developing countries<br>Advanced countries<br>Mixed sample |
| Exchange rate level | de Carvalho Filho (2010) (-)<br>Rose (2014) (x)<br>Lane and Van Den Heuvel (1998) (x) | Mixed sample<br>Mixed sample<br>UK |
| Exchange rate volatility | Almeida and Goodhart (1998) (x)<br>Ardakani et al. (2018) (+)<br>Ardakani et al. (2018) (x)<br>Batini and Laxton (2007) (-)<br>Berganza and Broto (2012) (+)<br>Edwards (2007) (x)<br>IMF (-)<br>Kuttner and Posen (2001) (x)<br>Lin (2010) (+)<br>Lin (2010) (-)<br>Ouyang and Rajan (2016) (x)<br>Ouyang et al. (2016) (+)<br>Ouyang et al. (2016) (x)<br>Pontines (2013) (-)<br>Pontines (2013) (+) | Advanced countries<br>Advanced countries<br>Developing countries<br>EMEs<br>EMEs<br>Mixed sample<br>EMEs<br>Mixed sample<br>Advanced countries<br>EMEs<br>EMEs<br>Advanced countries<br>Developing countries<br>Developing countries<br>Advanced countries |
| | Rose (2007) (x) | Mixed sample |
| Current account | Lin (2010) (x)<br>Rose (2014) (x) | Mixed sample<br>Mixed sample |
| Budget balance | Abo-Zaid and Tuzemen (2012) (x)<br>Abo-Zaid and Tuzemen (2012) (+)<br>Combes et al. (2014) (+)<br>Minea and Tapsoba (2014) (x)<br>Minea and Tapsoba (2014) (+)<br>Minea and Tapsoba (2014) (+)<br>Rose (2014) (x) | Advanced countries<br>Developing countries<br>Mixed sample<br>Advanced countries<br>Mixed sample<br>Developing countries<br>Mixed sample |
| Public revenue | Lucotte (2012) (+)<br>Minea et al. (2021) (+)<br>Fry-McKibbin and Wang (2014) (-)<br>Miles (2007) (-) | EMEs<br>Developing countries<br>Mixed sample<br>EMEs |
| Government debt | Ardakani et al. (2018) (-)<br>Fry-McKibbin and Wang (2014) (-)<br>Fry-McKibbin and Wang (2014) (x)<br>Rose (2014) (x) | Mixed sample<br>Advanced countries<br>EMEs<br>Mixed sample |
| Unemployment rate | de Carvalho Filho (2010) (x) | Mixed sample |

Note: "+" indicates that IT is associated with higher value of a macroeconomic variable (e.g., inflation, output, debt etc.); "-" indicates that IT is associated with lower value of a macroeconomic variable; "x" indicates statistically insignificant, economically negligible or mixed (non-robust) evidence.

Table 4. Empirical studies on the effects of IT on sacrifice ratios

| Study | Data and methodology | Main findings |
|---|---|---|
| Almeida and Goodhart (1998) | 13 OECD countries during 1981-1997; simple comparison between sacrifice ratios | (x) |
| Ardakani et al. (2018) | 98 advanced and developing countries during 1998-2013; propensity score matching | (-) in advanced countries<br>(x) in developing countries |
| Brito (2010) | 24 OECD countries during 1990-2005; OLS | (x) |
| Brito and Bystedt (2010) | 46 developing countries during 1980-2006; GMM | (x) |
| Chortareas et al. (2003) | 21 OECD countries during 1990-2000, SUR | (x) |
| Gonçalves and Carvalho (2008) | 40 OECD and developing countries, 1980-2006; OLS | (-) |
| Gonçalves and Carvalho (2009) | 30 OECD, 1980-2006; OLS, Heckman's two-stage procedure | (-) |
| Corbo et al. (2002) | 23 advanced and developing countries during 1980-2000; simple comparison | (x) |
| Debelle (1996) | NZ, Australia, and Canada during 1974-1993; simple comparison | (+) |
| Laubach and Posen (1997) | 8 advanced countries during 1971-1993; OLS | (+) |
| Magkonis and Zekente (2020) | 42 OECD and non-OECD countries during 1975-2015; Bayesian model averaging | (x) in both samples |
| Mazumder (2014) | 189 advanced and developing countries during 1972-2007; OLS regression and fixed-effects model | (x) in both OECD and developing countries<br>(+) in high-income countries<br>(x) in middle-income and low-income countries |
| Roux and Hofstetter (2014) | OECD countries during 1990-2006; OLS | (-) |
| Sethi and Acharya (2019) | 13 Asian countries during 1970-2014; OLS | (-) |
| Stojanovikj and Petrevski (2020) | 44 EMEs during 1970-2017; OLS | (+) |
| Tunali (2008) | 53 OECD and developing countries during 1990s-2007; 2SLS | (x) in both samples |

Note: "+" indicates that IT increases the sacrifice ratio; "-" indicates that IT reduces the sacrifice ratio; "x" indicates either mixed evidence or statistically insignificant results.